\documentclass{article}

\usepackage{arxiv}

\usepackage[utf8]{inputenc} % allow utf-8 input
\usepackage[T1]{fontenc}    % use 8-bit T1 fonts
\usepackage{hyperref}       % hyperlinks
\usepackage{url}            % simple URL typesetting
\usepackage{booktabs}       % professional-quality tables
\usepackage{amsfonts}       % blackboard math symbols
\usepackage{amsmath}
\usepackage{graphicx}
\graphicspath{ {./images/} }
\usepackage{cleveref}
\usepackage{natbib}
\usepackage{float}
\usepackage{upgreek} 
\usepackage{chemmacros}
\usepackage{chemformula}
\usepackage{makecell}
\usepackage{longtable}
\usepackage{float}
\usepackage[section]{placeins}
\usepackage{lscape}
\usepackage[locale=US,group-separator={,},separate-uncertainty=true,multi-part-units=single,per-mode=reciprocal]{siunitx}
\DeclareSIUnit\gCDW{g\textsubscript{CDW}}

\newcommand{\cmfa}{\textsuperscript{13}C-MFA}
\newcommand{\CG}{\textit{C.~glutamicum}}
\newcommand{\CGcap}{{\rmfamily\bfseries\itshape C.\ glutamicum}}

\newcommand{\cflux}{\textsc{13CFLUX}}
\DeclareMathOperator*{\argmin}{arg\,min}

\title{Automated multi-dataset INST \textsuperscript{13}C metabolic flux analysis at microliter scale reveals robust fluxes but variable metabolite pools in \textit{Corynebacterium glutamicum}}

\author{
    Jochen Nie{\ss}er\textsuperscript{1,2,$\dagger$}, 
    Anton Stratmann\textsuperscript{1,3,$\dagger$}, 
    Martin Bey\ss\textsuperscript{1,3}, 
    Wolfgang Wiechert\textsuperscript{1,3}, \\ 
    \textbf{Katharina N{\"o}h\textsuperscript{1,$\ddagger$,$\ast$}, 
    Stephan Noack\textsuperscript{1,$\ddagger$,$\ast$}} \\[2ex]
    \textsuperscript{1}Institute of Bio- and Geosciences, IBG-1: Biotechnology, Forschungszentrum Jülich, Jülich, Germany \\
    \textsuperscript{2}Institute of Biotechnology, RWTH Aachen University, Aachen, Germany \\
    \textsuperscript{2}Computational Systems Biotechnology (AVT.CSB), RWTH Aachen University, Aachen, Germany \\[2ex]
    \textsuperscript{$\dagger, \ddagger$} {Equal contributions}\\[2ex]
    \textsuperscript{$\ast$} {Corresponding authors \href{email:k.noeh@fz-juelich.de}{k.noeh@fz-juelich.de}, \href{email:s.noack@fz-juelich.de}{s.noack@fz-juelich.de}}
}

\begin{document}

\maketitle

\begin{abstract}
Isotopically non-stationary metabolic flux analysis (INST \cmfa) provides unique insights into cellular physiology but is typically limited by low throughput and high experimental costs. Here, we present a miniaturized and automated workflow that integrates transient isotope labeling experiments with advanced computational modeling to enable parallel INST \cmfa{} at microliter scale. The approach is demonstrated for an evolved \textit{Corynebacterium glutamicum} strain capable of efficient growth on ethanol, a substrate for which isotopically stationary \cmfa{} is inherently limited due to low labeling diversity. Using robotic liquid handling, rapid hot isopropanol quenching, and LC-QToF-MS-based analytics, highly informative datasets were generated from parallel 48-well experiments with different ethanol tracers. Multi-dataset INST \cmfa{} unlocked joint estimation of intracellular fluxes and metabolite pool sizes and significantly improved flux precision compared to single-dataset analyses. While net fluxes were robust across datasets, pool size estimates exhibited variability and did not converge under joint inference, highlighting a fundamental methodological difference to single-dataset INST \cmfa{}. 
The resulting multi-dataset flux map reveals a central role of the glyoxylate shunt during growth on ethanol, consistent with metabolic adaption to C2-based substrate utilization. Overall, this work demonstrates that automated multi-dataset INST \cmfa{} is technically feasible and provides high-quality flux analysis at a fraction of the cost of conventional lab-scale bioreactor-based approaches. The presented workflow establishes a scalable framework for high-throughput quantitative fluxomics in microbial biotechnology and supports integration into iterative strain engineering and biofoundry pipelines.
\end{abstract}


%%%%%%%%%%%%%%%%%%%%%%%%%%%%%%%%%%%%%%%%%%%%%%%%%%%%%%%%%%%%%%%%%%%%%%%%%%%%%%%%%%%%%%%%%%%%%%%%%%%%
\section*{INTRODUCTION}
\label{sec:introduction}
%%%%%%%%%%%%%%%%%%%%%%%%%%%%%%%%%%%%%%%%%%%%%%%%%%%%%%%%%%%%%%%%%%%%%%%%%%%%%%%%%%%%%%%%%%%%%%%%%%%%

Metabolic reaction rates (fluxes) provide the most direct functional readout of cellular phenotypes, integrating regulatory effects across multiple omics layers, including gene expression, translation, and post-translational protein modifications, into a quantitative description of metabolism~\citep{Sauer2006, Nielsen2003}. Their quantification under metabolic (pseudo-)steady conditions is enabled by metabolic flux analysis (MFA), which combines metabolic network models with extracellular rate measurements and, in the case of \cmfa{}, isotopic labeling data to infer intracellular flux distributions~\citep{Niedenfuhr2015}. In isotope labeling experiments (ILE), labeled substrates are administered and fractional labeling enrichments in intracellular metabolites and/or proteins are measured using mass spectrometry (MS) or nuclear magnetic resonance (NMR). By integrating the observed labeling patterns and extracellular rates in a metabolic network model, fluxes are estimated via iterative fitting.

Two main \cmfa{} approaches exist: (1) Isotopically stationary (IST) \cmfa{} relies on steady-state labeling data and, in advanced implementations such as COMPLETE-MFA, multiple tracers are integrated to improve flux resolution~\citep{Leighty2013}. (2) Isotopically non-stationary (INST) \cmfa{} exploits time-resolved labeling dynamics, enabling simultaneous estimation of fluxes and intracellular metabolite pool sizes (concentrations), reducing labeling times, and providing additional information for model validation~\citep{Noh2007, Noh2011}.
A fundamental limitation of IST \cmfa{} arises for substrates with a low number of carbon atoms, such as \ch{CO2}, methanol, or ethanol, all central to renewable feedstocks in emerging \ch{CO2}-based bioeconomy concepts~\citep{Bachleitner2023}, as they yield insufficient labeling diversity at isotopic steady state. In these cases, INST \cmfa{} becomes essential. 

Despite these methodological advances, the design and execution of informative ILEs remain constrained by cost, automation, and scalability. Isotopically labeled substrates represent the dominant cost factor~\citep{Noh2018}, scaling with both cultivation volume and labeling duration, thereby limiting ILE throughput. Advanced bioreactor concepts, such as sensor reactors (with 1~L working volume)~\citep{ElMassaoudi2003, Drysch2003} and parallel mini-bioreactor platforms with a working volume of 50~mL~\citep{Heux2014, Fina2023}, have reduced volumes and thereby labeling costs. Simultaneously, these setups have enabled sampling automation for improved standardization. However they do not provide a viable option for high-throughput INST \cmfa, as they either scale purely with the number of ILEs or remain restricted to IST-based protocols.

Recently, the emergence of biofoundries capable of generating large strain libraries has created a growing demand for high-throughput fluxomics~\citep{Rosch2024}. Miniaturized, fully roboterized cultivation platforms offer a promising route to address experimental limitations by limiting tracer cost and enabling parallel and fully automated experimentation~\citep{Unthan2015, Hemmerich2019, Niesser2022}. While these developments paved the way for INST ILEs, they also introduce a second bottleneck: the computational analysis of large-scale, time-resolved ILE data sets. INST \cmfa{} requires solving complex, nonlinear inverse problems, and the joint evaluation of multiple datasets further increases computational complexity. Recent development of high-performance \cmfa{} software, especially \texttt{13CFLUX}, has significantly improved simulation performance, scalability, and design capability~\citep{Stratmann2025}. Yet, integrated workflows combining automated experimentation with scalable, multi-dataset evaluations following the COMPLETE-MFA concept to INST \cmfa{} have not been attempted.

Here, we address this gap by integrating automated microliter-scale experimentation with a computational workflow for multi-dataset INST \cmfa{}, spanning experimental design (ED), ILE execution, data processing, and model-based flux inference within an iterative cycle. As a case study, we investigate the industrial workhorse \CG{}, widely used for the production of value-added compounds from renewable feedstocks~\citep{Liu2025, Marienhagen2025, Kurpejovic2025}. An evolved strain capable of efficient growth on ethanol (WT\_ETH-evo) provides a unique opportunity to dissect the underlying metabolic adaptations~\citep{Halle2023}.

Resolving intracellular fluxes in this system is particularly challenging, as the two-carbon structure of ethanol restricts labeling diversity in central carbon metabolism, limiting the informativeness of IST \cmfa{} and necessitating dynamic INST approaches. This combination of experimental constraints and industrial relevance establishes \CG{} WT\_ETH-evo as a stringent test case to demonstrate how automated, multi-dataset INST \cmfa{} workflows can overcome fundamental limitations of flux analysis in biofoundry-driven strain development.

%%%%%%%%%%%%%%%%%%%%%%%%%%%%%%%%%%%%%%%%%%%%%%%%%%%%%%%%%%%%%%%%%%%%%%%%%%%%%%%%%%%%%%%%%%%%%%%%%%%%
\section*{METHODS}
\label{sec:methods}
%%%%%%%%%%%%%%%%%%%%%%%%%%%%%%%%%%%%%%%%%%%%%%%%%%%%%%%%%%%%%%%%%%%%%%%%%%%%%%%%%%%%%%%%%%%%%%%%%%%%
\subsection*{Automated transient ILEs}
\label{subsec:transILEs}
%%%%%%%%%%%%%%%%%%%%%%%%%%%%%%%%%%%%%%%%%%%%%%%%%%%%%%%%%%%%%%%%%%%%%%%%%%%%%%%%%%%%%%%%%%%%%%%%%%%%
The presented experiments have been conducted with the strain \CG{} WT\_EtOH-Evo which has been derived from \CG{} ATCC~13032 (WT) in an adaptive laboratory evolution experiment \citep{Halle2023}. The strain was grown in CGXII medium \citep{keilhauer1993} composed of \SI{42}{g.L^{-1}} 3-(N-morpholino)propanesulfonic acid (MOPS) buffer, \SI{5}{g.L^{-1}} urea, \SI{20} {g.L^{-1}} ammonium sulfate, \SI{1}{g.L^{-1}} \ch{KH2PO4}, \SI{1}{g.L^{-1}} \ch{K2HPO4}, \SI{13.25}{mg.L^{-1}} \ch{CaCl2 . 2 H2O}, \SI{0.25}{g.L^{-1}} \ch{MgSO4 . 7 H2O}, \SI{10}{mg.L^{-1}} \ch{FeSO4 . 7 H2O}, \SI{10}{mg.L^{-1}} \ch{MnSO4 . H2O}, \SI{0.02}{mg.L^{-1}} \ch{NiCl2 . 6 H2O}, \SI{0.313}{mg.L^{-1}} \ch{CuSO4 . 5 H2O}, \SI{1}{mg.L^{-1}} \ch{ZnSO4 . 7 H2O}, \SI{0.2}{mg.L^{-1}} biotin, \SI{30}{mg.L^{-1}} protocatechuic acid with \SI{20}{g.L^{-1}} glucose as a carbon source for pre-cultures and \SI{1}{\%.(v.v^{-1})} ROTIPURAN Ethanol $\geq$ \SI{99.8}{\%} p.a. (Carl Roth GmbH + Co. KG, Karlsruhe, Germany) for main cultures.

Overnight pre-cultures were performed in baffled \SI{500}{mL} shaking flasks with \SI{10}{\%} filling volume at \SI{250}{rpm} and \SI{30}{\celsius} which were inoculated directly from cryo stocks. Samples from a pre-culture were centrifuged for \SI{5}{min} at \SI{4}{\celsius} and \SI{7000}{g}, washed once with phosphate buffer saline, and centrifuged again before resuspension of the cell pellet in main culture medium for inoculation with an optical density at a wavelength of \SI{600}{nm} (OD\textsubscript{600}) of \SI{0.5}{}. 

The robotic platforms (Mini Pilot Plants) used in this work all consisted of a Tecan Freedom Evo 200 liquid handler (Tecan Deutschland GmbH, Crailsheim, Germany) equipped with a liquid handling arm with \SI{8}{} fixed steel tips with a Teflon coating and a robotic manipulator arm. They further interfaced with numerous third party devices forming a general framework granting the freedom to perform various kinds of biological experiments. For more technical details on the Mini Pilot Plants see~\citep{Unthan2015, Hemmerich2019}.

Main cultures were performed in a BioLector I microbioreactor (Beckman Coulter GmbH, Baesweiler, Germany) in 48-well FlowerPlates at \SI{30}{\celsius} and \SI{1400}{rpm} featuring online measurements of dissolved oxygen (DO), pH, and backscatter (BS) with a gain of \SI{20}{}. After starting main cultivations on unlabeled ROTIPURAN Ethanol $\geq$ \SI{99.8}{\%} p.a. (Carl Roth GmbH + Co. KG, Karlsruhe, Germany), either \SI{100}{\%} 1-\textsuperscript{13}C-ethanol (Cambridge Isotope Laboratories, Andover, MA 01810 USA; \SI{99}{\%} purity), \SI{100}{\%} 2-\textsuperscript{13}C-ethanol, or \SI{100}{\%} U-\textsuperscript{13}C-ethanol (Santa Cruz Biotechnology, Inc., Heidelberg, Germany; \SI{99}{\%} purity) were administered successively in a column-wise fashion into wells, which were subsequently automatically harvested with increasing delays and quenched using automated hot isopropanol quenching \citep{Niesser2022}. The quenched extracts were then centrifuged for \SI{5}{min} at \SI{4500}{rpm} and \SI{4}{\celsius} and the supernatants were transferred to \SI{1.5}{mL} Eppendorf tubes which were stored in a \SI{-20}{\celsius} freezer until mass spectrometry analyses. The accuracy and reproducibility of the automated quenching step and its analytical robustness were validated (see~\Cref{sifig:FigS1} in the Supplemental information).

%%%%%%%%%%%%%%%%%%%%%%%%%%%%%%%%%%%%%%%%%%%%%%%%%%%%%%%%%%%%%%%%%%%%%%%%%%%%%%%%%%%%%%%%%%%%%%%%%%%%
\subsection*{Mass spectrometry}
\label{subsec:LCMSMS}
%%%%%%%%%%%%%%%%%%%%%%%%%%%%%%%%%%%%%%%%%%%%%%%%%%%%%%%%%%%%%%%%%%%%%%%%%%%%%%%%%%%%%%%%%%%%%%%%%%%%
All mass spectrometry analyses were conducted with an Agilent 1260 Infinity II HPLC system (Agilent Technologies, Waldbronn, Germany) connected to a Sciex TripleTOF6600 QqTOF device (AB Sciex Germany GmbH, Darmstadt, Germany) equipped with a Turbo V ion source. A previously validated method featuring ion-exchange chromatography with a \SI{150}{}~x~\SI{2}{mm} Phenomenex Luna SCX column (Phenomenex Ltd., Aschaffenburg, Germany) with a pore size of \SI{100}{\angstrom} and a particle size of \SI{5}{\micro m} preceded by a \SI{4}{}~x~\SI{2}{mm} SCX Security Guard cartridge (Phenomenex Ltd., Aschaffenburg, Germany) was used to analyze the labeling states of free amino acids \citep{reiter2021}. Samples injected at a volume of \SI{5}{\micro L} were separated with a gradient elution with \SI{5}{\%.(v.v^{-1})} acetic acid solution (A) and a \SI{15}{mM} ammonium sulfate solution adjusted to pH~6 with \SI{100}{\%} acetic acid (B) at a flow rate of \SI{0.4}{mL.min^{-1}} and a temperature of \SI{60}{\celsius}. The gradient was defined as follows: \SI{15}{\%} B at \SI{0}{min}, \SI{15}{\%} B at \SI{10}{min}, \SI{100}{\%} B at \SI{16}{min}, \SI{100}{\%} B at \SI{28}{min}, \SI{15}{\%} B at \SI{30}{min}, and \SI{15}{\%} B at \SI{35}{min}.

Peak data generated with the TripleTOF6600 QqTOF device was loaded into the vendor software Sciex MultiQuant (version 3.0.3) with a quantitation method specifying mass traces and their pertaining m/z ranges. Peak recognition and integration was performed with the MQ4 algorithm with default settings. Subsequently, results were reviewed visually and if necessary the baseline was manually adjusted, false negative or mislabeled peaks were corrected manually, false positives were de-selected to remove low-quality signals. Upon completion of these checks, the data were saved as a comma-separated value (csv) file. The calculation of tandem mass isotopomer distributions (TMID) was conducted with a Python script enabling standardized data processing. TMIDs were corrected for the abundance of naturally occurring heavy isotopes with an in-house tool, that processes the dataset using \texttt{IsoCorrectoR}~\citep{Heinrich2018} and validates the result independently using \texttt{ICT}~\citep{Jungreuthmayer2015}.  Finally, the quality of the labeling trajectories was visually judged and checked for consistency (e.g.~\Cref{sifig:FigS3} in the Supplemental information). Processed TMID data for all ILEs are available on GitHub (\url{https://github.com/JuBiotech/Supplement-to-Niesser-and-Stratmann-et-al.-Trends-Biotechnol.-2026}).

%%%%%%%%%%%%%%%%%%%%%%%%%%%%%%%%%%%%%%%%%%%%%%%%%%%%%%%%%%%%%%%%%%%%%%%%%%%%%%%%%%%%%%%%%%%%%%%%%%%%
\subsection*{Estimation of extracellular rates via bioprocess modeling}
\label{subsec:bioprocess}
%%%%%%%%%%%%%%%%%%%%%%%%%%%%%%%%%%%%%%%%%%%%%%%%%%%%%%%%%%%%%%%%%%%%%%%%%%%%%%%%%%%%%%%%%%%%%%%%%%%%
Data from the ILE with \CG{} WT\_EtOH-Evo on 1-\textsuperscript{13}C-ethanol were used to estimate the extracellular rates. Due to equal cultivation conditions, the rates of the two subsequent experiments on 2- and U-\textsuperscript{13}C-ethanol were assumed identical.

The rates were estimated from backscatter data, which were generated by the BioLector and pre-processed using the \texttt{bletl} tool~\citep{Osthege2022}. The data were analyzed by a bioprocess model based on Monod kinetics and parameterized with biomass growth and ethanol uptake:
\begin{align}
    \frac{dX}{dt} &= \mu \cdot X &\text{with} \quad \mu &= \mu_{\mathrm{max}} \cdot \frac{\mathrm{EtOH}}{K_{\mathrm{EtOH}} + \mathrm{EtOH}} & \text{and} \quad X(t_0) &= X_0\\   
    \frac{d\mathrm{EtOH}}{dt} &= q_{\mathrm{EtOH}} \cdot X &\text{with} \quad q_{\mathrm{EtOH}} &= -\frac{\mu}{Y_{X/\mathrm{EtOH}}} & \text{and} \quad \mathrm{EtOH}(t_0) &= \mathrm{EtOH}_0     
\end{align}

For model implementation and validation, \textit{OpenModelica}~\citep{Fritzson2020} in combination with the in-house Python package \texttt{estim8}~\citep{Latour2025} was used. The complete dataset with all \SI{24}{} batches was used for fitting the model parameters by applying the following replicate handling procedure: The maximum specific growth rate $\mu_{max}$ and the initial biomass concentration $X_0$ were allowed to vary between the replicates (local parameter), while the affinity constant $K_{\text{EtOH}}$ and the yield coefficient $Y_{X/\text{EtOH}}$ were assumed to be strain or process specific (global parameter). 

To enable mapping of backscatter observations to biomass variables (given as cell dry weight, CDW) an additional calibration experiment was performed. Here \CG{} WT\_EtOH-Evo was cultivated under equal environmental conditions (defined media with unlabeled ethanol as sole carbon and energy source) and technical setup (BioLector I and gain \SI{20}{}). A linear calibration model was fitted to the data within a predefined linear dynamic range as follows: $\text{BS} = 13.4254\cdot \text{CDW} + 15.3524$, and this function was added to the bioprocess model. Details on the calibration and bioprocess model as well as fitting procedures are found in the corresponding Jupyter notebook provided on GitHub 
(\href{https://github.com/JuBiotech/Supplement-to-Niesser-and-Stratmann-et-al.-Trends-Biotechnol.-2026}{https://github.com/JuBiotech/Supplement-to-Niesser-and-Stratmann-et-al.-Trends-Biotechnol.-2026}).

Finally, the resulting simulated trajectories of specific growth rates and ethanol uptake rates at the immediate time point before the labeling pulse were evaluated and metabolic stationarity was confirmed. The means and standard deviations of each rate were estimated from the \SI{24}{} single point estimates.

%%%%%%%%%%%%%%%%%%%%%%%%%%%%%%%%%%%%%%%%%%%%%%%%%%%%%%%%%%%%%%%%%%%%%%%%%%%%%%%%%%%%%%%%%%%%%%%%%%%%
\subsection*{Metabolic modeling}
\label{subsec:metabolic_model}
%%%%%%%%%%%%%%%%%%%%%%%%%%%%%%%%%%%%%%%%%%%%%%%%%%%%%%%%%%%%%%%%%%%%%%%%%%%%%%%%%%%%%%%%%%%%%%%%%%%%
The \textsuperscript{13}C metabolic network model of \CG{} was adapted from a reference model by \citet{Kappelmann2016} with modifications to reflect the specific strain characteristics and experimental conditions applied in this study. For instance, glucose uptake removed, and ethanol uptake as well as the glyoxylate shunt were added. A lumped biomass formation reaction was formulated based on literature-derived estimates of \CG{}'s cellular composition ~\citep{Eggeling2005, Kjeldsen2009}. The resulting metabolic model covers all central carbon metabolism (CCM) pathways and simplified routes for amino acid biosynthesis.
In total, it comprises \SI{70}{} balanced intracellular metabolites and \SI{80}{} intracellular reactions, of which \SI{22}{} are bidirectional and \SI{58}{} unidirectional. This results in \SI{29}{} free flux parameters (\SI{7}{} net and \SI{22}{} exchange fluxes), and \SI{56}{} pool size parameters, totaling \SI{85}{} parameters to be determined.
Model construction and visualization  were performed using \texttt{Omix} (ver.~2.1.2)~\citep{Droste2013}, including the specification of carbon atom transitions, tracers, and measurement configurations. The model was exported in FluxML format~\citep{Beyss2019}. All \texttt{Omix} and FluxML files used in this study are available on GitHub 
\href{https://github.com/JuBiotech/Supplement-to-Niesser-and-Stratmann-et-al.-Trends-Biotechnol.-2026}{https://github.com/JuBiotech/Supplement-to-Niesser-and-Stratmann-et-al.-Trends-Biotechnol.-2026}).

%%%%%%%%%%%%%%%%%%%%%%%%%%%%%%%%%%%%%%%%%%%%%%%%%%%%%%%%%%%%%%%%%%%%%%%%%%%%%%%%%%%%%%%%%%%%%%%%%%%%
\subsection*{Intracellular flux and pool size estimation}
\label{subsec:estimation}
%%%%%%%%%%%%%%%%%%%%%%%%%%%%%%%%%%%%%%%%%%%%%%%%%%%%%%%%%%%%%%%%%%%%%%%%%%%%%%%%%%%%%%%%%%%%%%%%%%%%
Intracellular fluxes $\mathbf{v}$ and metabolite pool sizes $\mathbf{X}$ of the \cmfa{} model were jointly estimated as part of the combined parameter vector $\boldsymbol{\theta}$. The parameters $\boldsymbol{\theta}$ are defined within a convex space $\mathcal{P}=\{ \boldsymbol{\theta}|\mathbf{S}\cdot\mathbf{v}=\boldsymbol{0}, \boldsymbol{\theta}_{\mathrm{lb}} \leq \boldsymbol{\theta} \leq \boldsymbol{\theta}_{\mathrm{ub}} \}$. The estimation was framed as a nonlinear regression problem, where the parameters $\boldsymbol{\theta}$ minimize the weighted sum of squared residuals (SSR) between simulated and observed data~\citep{Noh2006}
\begin{equation}
    \argmin_{\boldsymbol{\theta} \in \mathcal{P}}
    \sum_{i,j}\left(\frac{\mathbf{y}_{j}(t_i, \boldsymbol{\theta}) - \mathbf{y}_{j}^{\text{meas}}(t_i)}{\sigma_{y_j^{\text{meas}}(t_i)}}\right)^2 + 
    \sum_k \left(\frac{\mathbf{X}_{k} - \mathbf{X}_{k}^{\text{meas}}}{\sigma_{\mathbf{X}_k^{\text{meas}}}}\right)^2 +
    \sum_l \left(\frac{\mathbf{v}_{l} - \mathbf{v}_{l}^{\text{meas}}}{\sigma_{\mathbf{v}_l^{\text{meas}}}}\right)^2 \qquad
    \label{eqn:opt}
\end{equation}
Here, $\mathbf{y}^{\text{meas}} (t_i)$ represent the vector of observed TMIDs at time points $t_i$, $\mathbf{y}(t_i, \boldsymbol{\theta})$ the corresponding simulated TMIDs, and $\mathbf{\sigma}_{\mathbf{y}^{\text{meas}}(t_i)}$ their standard deviations (STD). Likewise, $\mathbf{X}^{\text{meas}}$ and $\mathbf{v}^{\text{meas}}$ represent the vectors of observed pool sizes (none in this study) and extracellular rate measurements (see~\nameref{subsec:bioprocess}), respectively.

To reduce the risk of being trapped in local minima when solving the nonlinear optimization problem (see~\Cref{eqn:opt}), a multi-start heuristic with \SI{1000}{} random initial points for $\boldsymbol{\theta}$ was applied. The best fit was determined to be the one with the lowest SSR. To quantify statistical parameter uncertainties, we applied linearized statistics based on the parameter covariance matrix $\mathbf{Cov}$, which was derived from the local curvature of the SSR surface at the best fit values. The \SI{95}{\%} confidence intervals (CI) were then calculated using the \num{0.975} quantile of the standard normal distribution~\citet{Theorell2017} with support restricted to the stoichiometrically feasible parameter space:
\begin{equation}
    \text{\textit{CI}}(\boldsymbol{\theta}_j) = \left.
    \left[\hat{\boldsymbol{\theta}}_j - q_{0.975} \cdot \sqrt{\mathbf{Cov}_{jj}}, \quad \hat{\boldsymbol{\theta}}_j + q_{0.975} \cdot \sqrt{\mathbf{Cov}_{jj}}\right] \right|_{\mathcal{P}}
    \label{eqn:ci}
\end{equation}
To categorize the estimated parameters qualitatively in terms of their identifiability, the coefficient of variation (CV)
\begin{equation}
    \text{\textit{CV}}(\boldsymbol{\theta}_j) = \frac{\sqrt{\mathbf{Cov}_{jj}}}{|\hat{\boldsymbol{\theta}}_j|}
\end{equation}
is used with CV $ \leq 0.25$ - well-identifiable, $0.25<$ CV $\leq 1$ - weakly identifiable, and CV $>1$ non-identifiable.

All simulations and parameter estimations were performed using the high-performance simulation engine \cflux{} (ver.~3.0.0, \citep{Stratmann2025}), using the interior-point optimizer \texttt{IPOPT} (ver.~3.14.14) for solving the regression problem in \Cref{eqn:opt}~\citep{Wachter2006}. Feasible random initial points for the parameters of the multi-start optimizations were generated using the highly optimized polytope sampling library \texttt{hopsy}~\citep{Paul2024}.

%%%%%%%%%%%%%%%%%%%%%%%%%%%%%%%%%%%%%%%%%%%%%%%%%%%%%%%%%%%%%%%%%%%%%%%%%%%%%%%%%%%%%%%%%%%%%%%%%%%%
\section*{RESULTS}
\label{sec:results}
%%%%%%%%%%%%%%%%%%%%%%%%%%%%%%%%%%%%%%%%%%%%%%%%%%%%%%%%%%%%%%%%%%%%%%%%%%%%%%%%%%%%%%%%%%%%%%%%%%%%
\subsection*{Workflow for microliter-scale INST \cmfa}
\label{subsec:workflow}
%%%%%%%%%%%%%%%%%%%%%%%%%%%%%%%%%%%%%%%%%%%%%%%%%%%%%%%%%%%%%%%%%%%%%%%%%%%%%%%%%%%%%%%%%%%%%%%%%%%%

Transferring INST \cmfa{} to automated microliter-scale platforms introduces both new challenges and opportunities for ILE execution and data analysis. While the required components exist in principle, their integration into a cohesive, high-throughput workflow remains non-trivial. A key opportunity lies in the use of microtiter plates, which enable the parallel execution of multiple INST ILEs within a single experimental run. This increased throughput enables the generation of complementary datasets that can be jointly analyzed, but simultaneously imposes new demands on computational pipelines for ILE design, data processing, and coherent multi-dataset integration.

\begin{figure}[htp!]
    \centering
    \includegraphics[width=\textwidth]{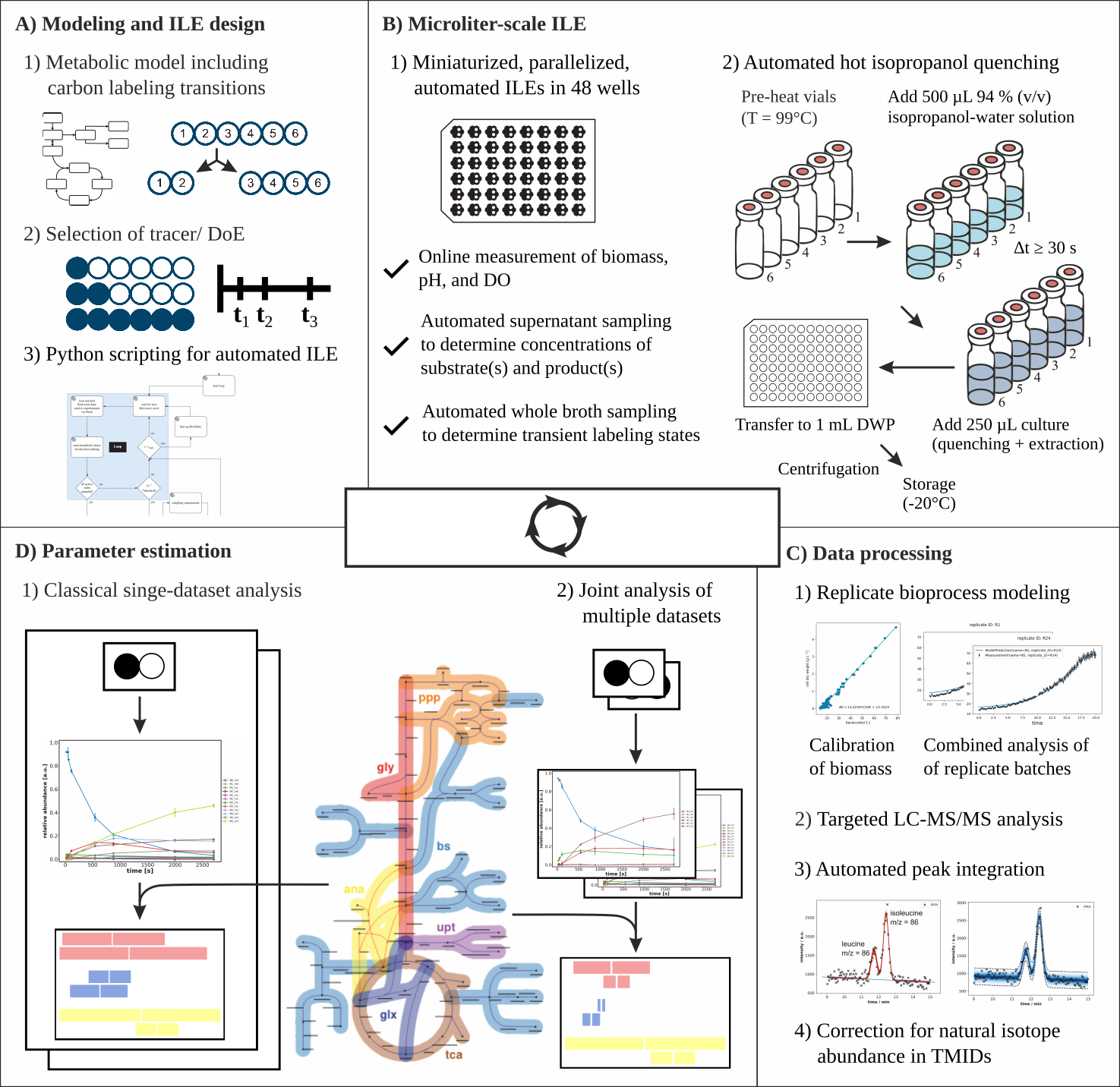}
    \caption{\textbf{Workflow for microliter-scale INST \cmfa.} 
    The workflow comprises four modules: (A) model formulation, ED, and ILE scripting; (B) automated execution of ILEs and sample processing; (C) data processing; and (D) model-based estimation of fluxes and pool sizes. 
    Miniaturized automation enables parallel execution of multiple INST ILEs, increasing throughput while reducing tracer cost. The resulting datasets can be evaluated individually or jointly, enabling consistency checks and improved information gain. The workflow is iterative and supports the targeted design of subsequent ILEs.
    }
\label{fig:figure-1}
\end{figure}

To address these challenges, we present an integrated workflow for multi-dataset INST \cmfa{}, spanning the full range from ED to flux estimation (see~\Cref{fig:figure-1}). The workflow is organized into four functional modules: 
(A) ILE design and preparation, 
(B) automated cultivation and time-resolved sampling, 
(C) mass spectrometric data acquisition and processing, and 
(D) computational model-based flux and pool size estimation. 
In the following, we highlight key aspects of each module in the context of microliter-scale, high-throughput INST \cmfa{}.

\begin{enumerate}

\item[A] 
    
    The first module addresses pre-ED of both the \textit{in silico} and wet-lab components of INST \cmfa{}. It begins with the formulation of a stoichiometric metabolic network model, in which reactions are annotated with reversibility and carbon atom transitions. Such models are adopted from literature or constructed using pathway databases such as KEGG~\citep{Kanehisa2000} and BioCyc~\citep{King2016}, supported by visualization and modeling tools including Escher~\citep{King2015} and Omix~\citep{Droste2013}, the latter enabling visual atom mapping and export to the standarized FluxML format~\citep{Beyss2019}.
    
    Building on the model, ED defines tracer selection and sampling schedules, guided by prior knowledge or assumed fluxes and pool sizes~\citep{Noh2007}. In microtiter-scale settings, additional design dimensions arise, including the number of biological replicates and the allocation of multiple ILEs within a single experimental run.
    
    These design choices are translated into executable experiments via automated liquid handling systems. An experimental control script (ECS) specifies all cultivation and sampling operations, including inoculation, labeling pulses, harvesting, and supernatant sampling for extracellular rate measurement~(M. Osthege, PhD thesis, RWTH Aachen University, 2023). The ECS is generated from standardized templates using a Python-based generator built on the \texttt{Jinja2} templating engine (\hypertarget{https://jinja.palletsprojects.com}{https://jinja.palletsprojects.com}), to ensure reproducibility while enabling rapid implementation of custom experimental ILE layouts.
    
\item[B] 

    While miniaturized cultivation platforms offer significant gains in automation, throughput, and cost efficiency, they impose constraints on experimental observability and introduce additional analytical challenges (see module C). In particular, microliter-scale ILEs are performed in batch mode, resulting in unknown residual substrate concentration at the time of tracer addition. In contrast to substrate-limited continuous bioreactors, this must be explicitly accounted for during model-based analysis (module D).

    A second implication of short labeling time windows is the presence of pre-existing unlabeled biomass components, which can distort the direct interpretation of labeling measurements. By focusing on free intracellular amino acids or CCM intermediates, the INST workflow avoids correction for this unlabeled fraction, thereby reducing an additional source of uncertainty.
    
    Automated ILEs provide online process data and supernatant measurements, enabling quantification of extracellular metabolites and by-products. Accurate capture of intracellular labeling dynamics under transient conditions critically depends on rapid and reliable quenching, as ongoing metabolic activity during sampling distorts labeling patterns. While cold methanol quenching followed by extraction in methanol–chloroform remains the gold-standard~\citep{Paczia2012, Tillack2012}, it is labor-intensive and incompatible with high-throughput workflows.
    
    To overcome this limitation, microliter-scale setups enable parallel execution of \SI{48}{} or more INST ILEs without time-critical manual intervention. Central to automation approach is a rapid hot isopropanol quenching protocol that terminates metabolic activity within seconds, enabling reliable analysis of isotope incorporation into free intracellular amino acids experiencing second-to-minute-scale resolution~\citep{Niesser2022}.

\item[C] 

    The third module transforms raw ILE data into inputs for flux and pool size estimation (see~\Cref{fig:figure-1}C) and comprises two processing pipelines: (i) estimation of extracellular rates from bioprocess data and (ii) extraction of intracellular labeling patterns from mass spectrometry measurements.

    At the process level, each batch cultivation is evaluated using a bioprocess model to estimate extracellular rates such as growth, substrate uptake, and product formation~\citep{noack2017}. Incorporating mechanistic knowledge supports validation of key \cmfa{} assumptions, including metabolic stationarity, and enables consistency checks such as carbon balance closure. Tools such as \texttt{PhysioFit}~\citep{LeGregam2024}, \texttt{pyFOOMB}~\citep{Hemmerich2021}, and \texttt{estim8}~\citep{Latour2025} facilitate this analysis, with the latter two particularly suited for transient ILEs due to explicit handling of tracer pulsing, replicates, and uncertainty propagation.

    In parallel, intracellular labeling patterns are quantified using LC-MS/MS or LC-QToF-MS~\citep{Kappelmann2017}. These measurements generate thousands of chromatographic peaks across metabolites, time points, and replicates, all of which must be accurately integrated to derive mass isotopomer distributions. While vendor software (e.g. \texttt{Sciex MultiQuant}) supports automated peak detection, manual inspection and correction remain common, constituting a major bottleneck for high-throughput workflows. The Python package \texttt{PeakPerformance}~\citep{Niesser2024} addresses this by automating peak fitting, model selection, and uncertainty quantification, enabling consistent processing across datasets; however, its integration into \cmfa{} pipelines remains limited by differences in statistical frameworks and computational scalability. 

    Following peak integration, time-resolved TMIDs are calculated, corrected for natural isotope abundance~\citep{Niedenfuhr2016}, and evaluated across biological replicates to obtain means and standard deviations. A key challenge in miniaturized INST workflows is the low signal-to-noise ratio of labeling data, particularly for positionally labeled tracers and early time points, which leads to discontinuous trajectories and increased uncertainty. Because these uncertainties directly affect parameter estimation (module D)~\citep{Wiechert2025}, their accurate assessment is critical: overestimation reduces the influence of measurements and inflates parameter CIs, whereas underestimation can lead to biased or overconfident flux estimates.

    Consequently, rigorous data quality assessment is essential prior to model fitting, including identification of low-intensity signals, outliers, and inconsistent replicate behavior, and, where necessary, exclusion or reweighting of unreliable measurements. Standardization and automation are prerequisites for high-throughput ILE workflows, which in turn enable robust characterization of measurement uncertainty through biological replicates.

\item[D] 

    The final module integrates the processed data into the \cmfa{} model defined in A, enabling joint estimation of intracellular fluxes and pool sizes from INST ILE datasets (\Cref{fig:figure-1}D).
    
    While multi-dataset analysis has become standard in IST \cmfa{} to improve flux resolution (e.g., \citep{Long2019}), INST \cmfa{} has so far been limited to single-dataset evaluations. Robotic microliter-scale experimentation now enables generation of multiple ILE datasets within a single run, thereby opening the door to multi-dataset INST \cmfa{}. However, this shift increases the complexity and computational demands of the evaluation workflow.
    
    Addressing these demands requires high-performance and flexible software infrastructures. Platforms such as \texttt{INCA}~\citep{Rahim2022}, \texttt{influx\_si}~\citep{Sokol2012}, and \cflux~\citep{Stratmann2025} support multi-dataset INST workflows, with \cflux{} offering superior performance in terms of simulation speed, accuracy, and scalability~\citep{Stratmann2025}. Its efficient implementation enables large-scale multi-start optimization, which is crucial for robust parameter estimation in \cmfa{}.

    Following parameter estimation, uncertainties of fluxes and pool sizes must be estimated. Linearized statistical methods provide fast but approximate estimates, whereas Monte Carlo, profile likelihood and Bayesian approaches improve the quality of uncertainty quantification at higher computational cost~\citep{Theorell2017}. Efficient and scalable computational pipelines are therefore critical for high-quality analyses. 

    Beyond performance, reproducibility and scriptability are key requirements for efficient evaluations. GUI-based tools such as \texttt{INCA} offer limited automation, whereas \cflux{} supports scripted pipelining and interaction with FluxML documents, allowing flexible integration of multiple datasets and version-controlled propagation of model or data updates.

    Finally, best-fit parameters and their CIs are visualized as flux maps aligned with the underlying model structure. Formats such as the Omix Visualization Language (OVL)~\citep{Droste2013} support this representation and allow interactive refinement. Thereby, changes introduced at this stage are directly propagated by re-running the scripted evaluation pipeline, while the resulting insights inform  subsequent ED, thereby closing the loop of the outlined INST \cmfa{} workflow.

\end{enumerate}

The closed-loop structure of the microliter-scale INST \cmfa{} workflow aligns naturally with the Design-Build-Test-Learn (DBTL) cycle, supporting iterative learning in knowledge-based strain engineering. While the workflow is conceptually organized as a linear sequence of four modules, its practical implementation relies on flexible data processing and analysis pipelines  within each module. As a result, execution is inherently adaptive and often iterative. Iterations occur both within and between modules, for instance when revising processed TMIDs (C) after model-based parameter estimation (D), or refining numerical settings following initial test runs. 

This requires composable pipelines tailored to experimental objectives, data characteristics, and computational constraints. These pipelines are integrated into a configurable workflow, whose effective operation depends on well-calibrated analytical instrumentation and adequate computational resources, both essential for scalable, high-throughput INST \cmfa{}.

%%%%%%%%%%%%%%%%%%%%%%%%%%%%%%%%%%%%%%%%%%%%%%%%%%%%%%%%%%%%%%%%%%%%%%%%%%%%%%%%%%%%%%%%%%%%%%%%%%%%
\subsection*{INST-ILE with \CG{} WT\_EtOH-Evo on 1-\textsuperscript{13}C labeled ethanol}
\label{subsec: R&D 1-13C EtOH}
%%%%%%%%%%%%%%%%%%%%%%%%%%%%%%%%%%%%%%%%%%%%%%%%%%%%%%%%%%%%%%%%%%%%%%%%%%%%%%%%%%%%%%%%%%%%%%%%%%%%

To demonstrate the microtiter-based automated INST \cmfa{} workflow, we applied it to quantify intracellular fluxes and pool sizes of the evolved \CG{} strain WT\_EtOH-Evo grown on ethanol as sole carbon source.
Due to the limited tracer options for ethanol and the absence of prior flux and pool size information, a classical model-based ED, as described previously~\citep{Noh2006}, was not feasible. Instead, sampling time points were heuristically selected based on prior data. In particular, the ethanol uptake of WT\_EtOH-Evo in lab-scale cultivations was reported to be roughly 1.5 times slower than glucose uptake (ethanol uptake rate \SI{16}{C-mmol.g_{X}^{-1}.h^{-1}} for WT\_ETH-Evo vs. glucose uptake rate for the WT \SI{27}{C-mmol.g_{X}^{-1}.h^{-1}})~\citep{Halle2023}, suggesting slower labeling dynamics in the Embden-Meyerhof-Parnas (EMP) and pentose phosphate (PP) pathways. Together with enrichment dynamics from a previous WT ILE using U-\textsuperscript{13}C glucose ~\citep{Niesser2022}, this informed the selection of seven time points covering the transient labeling regime (\SI{24}-\SI{1800}{s},see~\Cref{sitab:TabS1} in the Supplemental information).
Biological triplicates were generated at each time point, and 1-\textsuperscript{13}C ethanol was selected as a cost-efficient, yet informative positional tracer. In total, \SI{24}{} parallel batch cultivations were implemented, \SI{21}{} pulse-labeled batch cultures and \SI{3}{} pulse-free unlabeled controls to monitor the growth dynamics. 

All cultivations were executed via the automated ECS workflow generator, including labeling pulses, time-resolved sampling, and rapid hot isopropanol quenching (see~\nameref{subsec:transILEs}, \Cref{sifig:FigS2} in the Supplemental information). Bioprocess modeling of all cultivations (see~\nameref{subsec:bioprocess}) yielded a specific ethanol uptake rate of $q_S$ = \SI{7.95 \pm 0.68}{mmol.g_{X}^{-1}.h^{-1}} and a specific growth rate of $\mu_{max}$=\SI{0.19 \pm 0.02}{h^{-1}}, consistent with previous bioreactor-scale results~\citep{Halle2023}. The transient samples were analyzed using LC-QToF-MS to quantify intracellular amino acid labeling dynamics, yielding time-resolved mean TMIDs with associated standard deviations after natural abundance correction (see~\nameref{subsec:LCMSMS}).

\begin{figure}[htp!]
    \centering
    \includegraphics[width=1\textwidth]{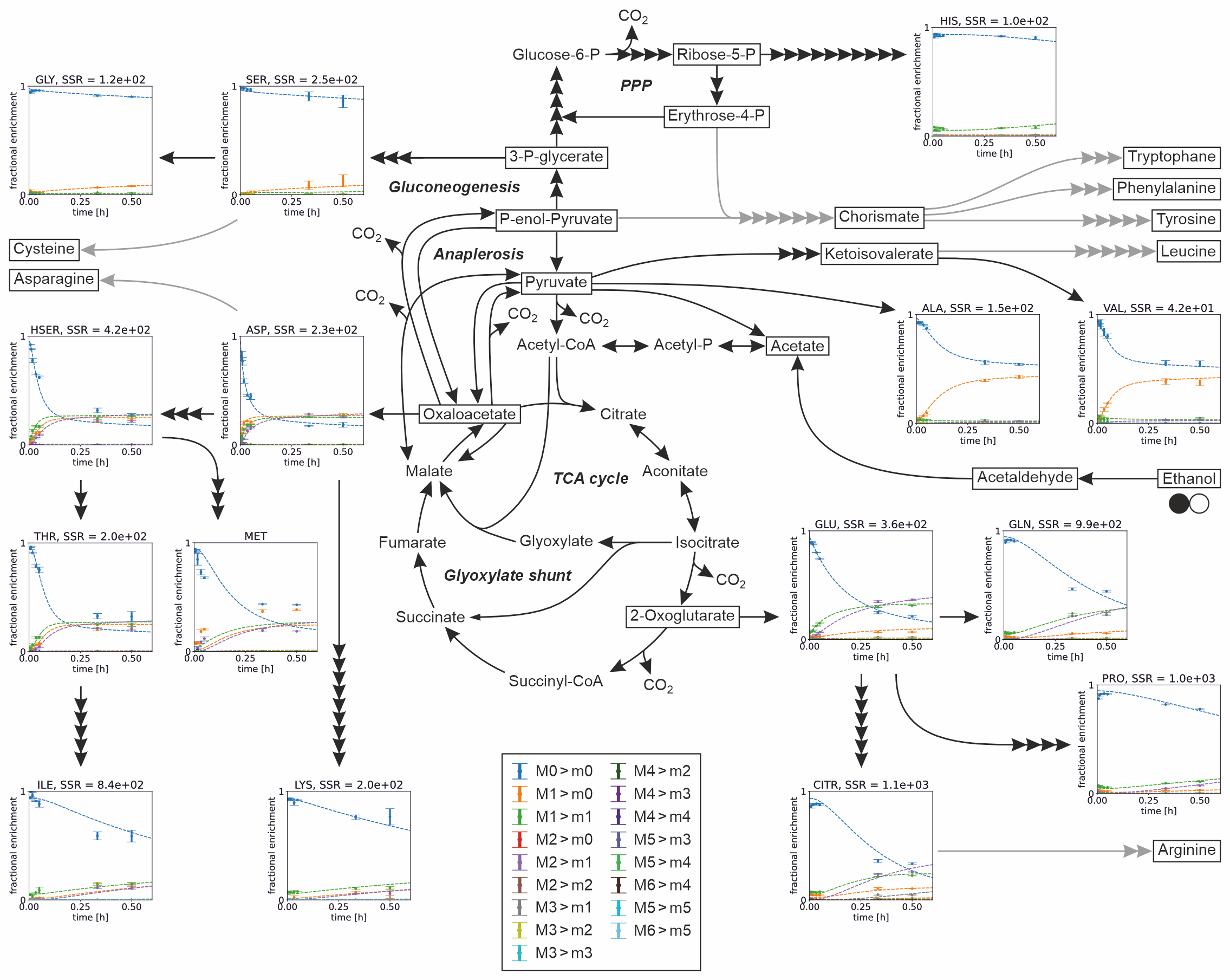}
    \caption{\textbf{Transient labeling dynamics of intracellular free amino acids during the 1-\textsuperscript{13}C ethanol ILE.} Data points originate from biological triplicates, and solid lines indicate simulated labeling trajectories at the best fit (see~\nameref{subsec:estimation}). The number of arrows corresponds to the number of reaction steps between metabolites. Grey arrows indicate amino acids for which no labeling measurements were available.}
    \label{fig:figure-2}
\end{figure}

The 1-\textsuperscript{13}C ethanol ILE revealed distinct labeling  dynamics across amino acids (see~\Cref{fig:figure-2}). Rapid incorporation within the first minute after the labeling pulse was observed for aspartate (ASP), homoserine (HSER), and threonine (THR), which approached isotopic steady state within \SI{30}{min}. Intermediate dynamics were found for TCA and pyruvate-derived amino acids (e.g. glutamate - GLU, glutamine - GLN, citrulline - CITR, alanine - ALA, valine - VAL), whereas amino acids associated with glucolytic/gluconeogenic and PPP-derived precursors, including serine (SER), glycine (GLY), and histidine (HIS), exhibited little to no label incorporation within the observed time frame. Label propagation slowed along the biosynthetic pathways, with several amino acids (e.g., isoleucine (ILEU), lysine (LYS), proline (PRO)) not reaching an isotopic steady state, likely due to buffering effects by large precursor pools. Amino acids with insufficient signal quality (tryptophan - TRP, phenylalanine - PHE, tyrosine - TYR, leucine - LEU, cysteine - CYS, and asparagine - ASN, ornithine - ORN) were excluded from further analyses (see~\Cref{sitab:TabS1} in the Supplemental information).

The observed LC-QToF-MS-derived labeling patterns were consistent with expected carbon transitions in central metabolism (see~\Cref{sifig:FigS3}, \Cref{sifig:FigS4} in the Supplemental information). For instance, SER labeling was restricted to specific mass isotopomers, reflecting the transfer of the labeled carbon through phosphoenolpyruvate (PEP), while PPP-derived amino acids such as HIS showed minimal enrichment due to delayed label propagation along their biosynthetic pathways.

The STDs of TMIDs spanned five orders of magnitude, ranging from $3 \cdot 10^{-6}$ (close to the LC-QToF-MS detection limit) to $2\cdot 10^{-1}$ (see~\Cref{sifig:FigS5} in the Supplemental information). While approximately \SI{42}{\%} of these values aligned with literature values~\citep{Noh2018} and \SI{8}{\%} exceeded these values ($> 2.5 \cdot 10^{-2}$), the remaining STDs were considerably lower ($ < 10^{-3}$). The low STDs were predominantly associated with low-intensity signals as a result of a low fractional enrichment, indicating potential overconfidence. To mitigate the bias in parameter estimation, a lower bound of $3 \cdot 10^{-3}$ was applied to all STDs (see~\Cref{sifig:FigS6} in the Supplemental information). In our study, pool sizes were not directly measured, and consequently had to be inferred from the data in addition to the fluxes.

The processed TMIDs and extracellular rates were incorporated into the \cmfa{} model (see~\nameref{subsec:metabolic_model}) and fitted using \cflux. Simulated labeling curves for the best fit matched experimental data well for most amino acids (see~\Cref{sifig:FigS7} in the Supplemental information). Deviations were observed primarily for GLU-derived amino acids (GLN, PRO, and CITR), particularly at early time points, and methionine (MET) exhibited both qualitative and quantitative mismatches and was consequently excluded from further analysis. 

\begin{figure}[htp!]
    \centering
    \includegraphics[width=1\textwidth]{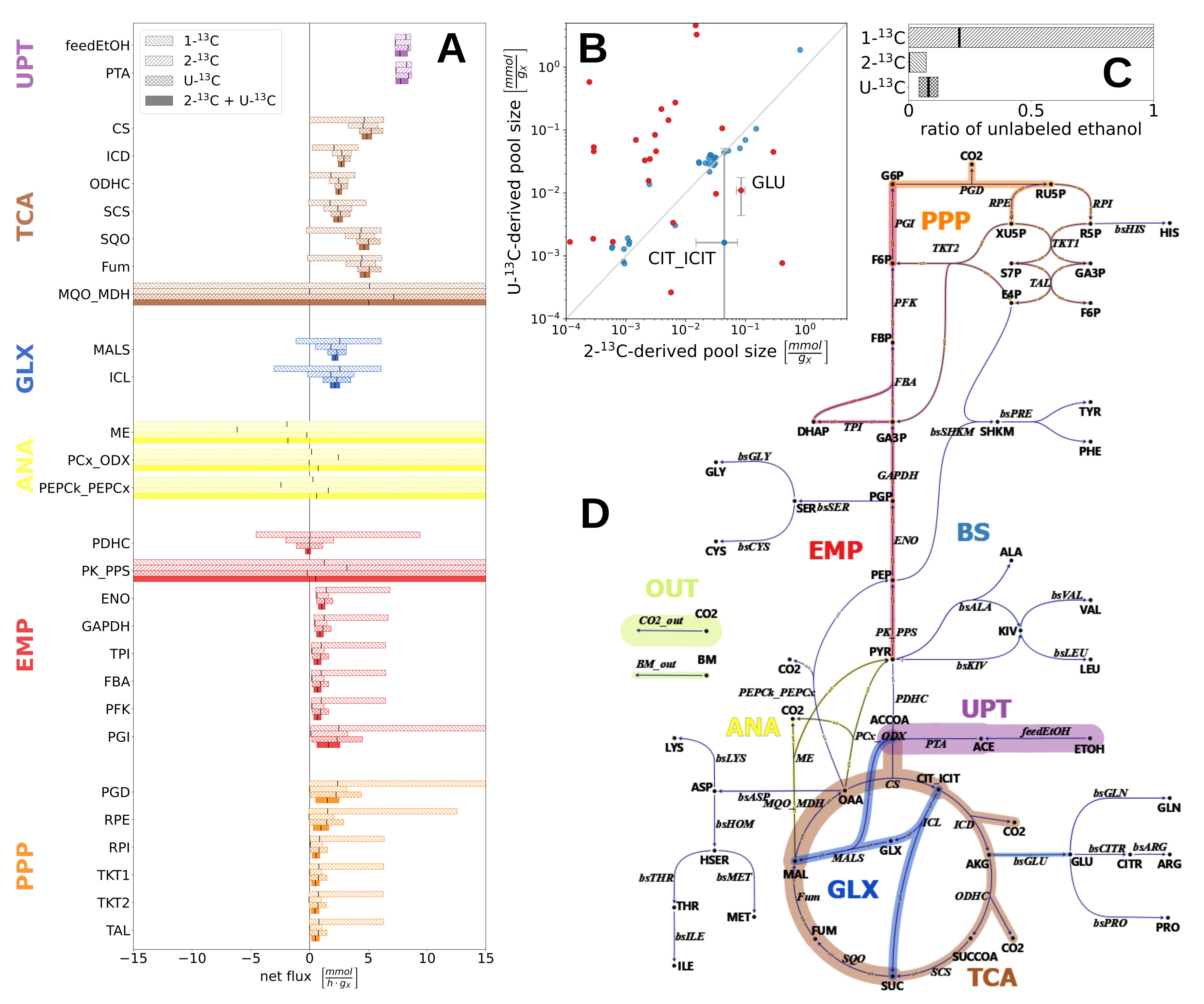}
    \caption{\textbf{Parameter estimates from single- and multi-dataset INST \cmfa{} analyses.}
    (A) Net flux estimates and corresponding CIs from individual and joint evaluations. Fluxes are colored by pathway as in (D).
    (B) Parity plot of identifiable pool size estimates for the 2- and U‑\textsuperscript{13}C ethanol ILEs. Each point represents one pool size. Red points indicate statistically different estimates when accounting for their CIs (e.g., GLU), whereas blue points denote consistent estimates (e.g.  CIT\_ICIT). 
    (C) Estimated fractions of residual (unlabeled) ethanol uptake in the three ILEs. A substantial fraction is observed for the U‑\textsuperscript{13}C ILE, but not for 1- and 2‑\textsuperscript{13}C ILEs.
    (D) Flux map derived from joint evaluation of the parallel INST datasets. Carbon from ethanol predominantly enters the TCA cycle, with an active GLX shunt and comparatively lower fluxes through EMP and PPP.}
    \label{fig:figure-3}
\end{figure}

Flux analysis revealed that the majority of ethanol-derived carbon is directed into the TCA cycle (see~\Cref{fig:figure-3}), with subsequent redistribution toward anaplerosis (ANA), gluconeogenesis (EMP), biosynthesis pathways, and CO\textsubscript{2} formation. A substantial flux entered the TCA cycle via citrate synthase (\texttt{CS}, \SI{4.65 +- 2.85}{mmol.g\textsubscript{X}^{-1}.h^{-1}}), with approximately half of this flux proceeding through the glyoxylate shunt (GLX, \SI{2.57 +- 1.90}{mmol.g_{X}^{-1}.h^{-1}}), underscoring its relevance during ethanol metabolism. In contrast, anaplerotic fluxes remained unresolved due to large CIs, rendering both their magnitudes and directions uncertain. This reflects known structural non-identifiability in this network region~\citep{Kappelmann2016}. This uncertainty also affected flux estimates in the EMP and PPP, such as for pyruvate dehydrogenase (\texttt{PDHC}) and phosphoenolpyruvate synthase (\texttt{PPS}). Compared to TCA and GLX fluxes, fluxes into EMP and PPP were comparatively small and associated with broad uncertainties.

A key source of uncertainty in the estimation arises from the absence of direct intracellular metabolite concentration measurements. Consequently, pool sizes had to be inferred solely from the labeling data, requiring their simultaneous estimation alongside fluxes. This increased the dimensionality of the problem and limits parameter identifiability, particularly for metabolite pools. In addition, uncertainty in the fraction of residual unlabeled ethanol in the medium introduces an additional degree of freedom in the model (see~\Cref{fig:figure-3}C).

Estimated intracellular pool sizes spanned three orders of magnitude, from $10^{-4}$~\si{mmol.g_{X}^{-1}} (e.g. ASP, HSER) up to $10^{-1}$~\si{mmol.g_{X}^{-1}} for GLU and AKG. Only a subset of the amino acid pool sizes were found to be reliably identifiable: the amino acid for which labeling dynamics was observed (i.e., ALA, CITR, GLN, GLU, HSER, ILE, PRO, THR) were determined with good and ASP with weak precision, besides the intermediates $\alpha$-ketoisovalerate (KIV) and diaminopimelate (DAP) (see~\Cref{sifig:FigS11} in the Supplemental information), whereas most other amino acid pool sizes remained practically non-identifiable. Despite these limitations, the 1‑\textsuperscript{13}C ethanol ILE yielded a high proportion of identifiable net fluxes. Based on the CV criterion (see~\nameref{subsec:estimation}), approximately \num{81}{\%} and \num{20}{\%} of the net fluxes and pool sizes were resolved, respectively, supporting quantitative conclusions on central carbon distribution.

%%%%%%%%%%%%%%%%%%%%%%%%%%%%%%%%%%%%%%%%%%%%%%%%%%%%%%%%%%%%%%%%%%%%%%%%%%%%%%%%%%%%%%%%%%%%%%%%%%%%
\subsection*{From single- to multi-dataset INST \textsuperscript{13}C-MFA}
\label{subsec:multi-dataset INST-13C-MFA}
%%%%%%%%%%%%%%%%%%%%%%%%%%%%%%%%%%%%%%%%%%%%%%%%%%%%%%%%%%%%%%%%%%%%%%%%%%%%%%%%%%%%%%%%%%%%%%%%%%%%

To improve flux and pool size precision beyond the 1‑\textsuperscript{13}C ethanol ILE, a second experimental cycle was performed. The insights gained so far allowed to perform a refined ED which was directed to transition from single- to multi-dataset INST \cmfa{}. While combining multiple ILE datasets with complementary tracers is well established in IST \cmfa{} to improve flux resolution, its implications for INST \cmfa{} have not yet explored.

Leveraging the high-throughput capacity of microtiter plates, a second experiment was designed to enable the parallel execution of two INST ILEs with complementary tracers while fully utilizing the \SI{48}{}-well format. To ensure statistical robustness, biological replicates were maintained, which constrained the design to eight sampling points per ILE. Tracer selection and sampling times were optimized using model-based ED (covariance-based A-criterion~\citep{Noh2006}). Positionally labeled 2-\textsuperscript{13}C ethanol and fully labeled U-\textsuperscript{13}C ethanol were selected to maximize labeling information, while the resulting non-uniform sampling schedule (eight sampling time points, see~\Cref{sitab:TabS1} in the Supplemental information) covered the transient labeling regime across observable amino acids. In both ILE settings, the design predicted a substantial reduction in parameter uncertainty (in terms of expected CIs of identifiable parameters), primarily driven by improved estimation of the labeled-to-unlabeled ethanol uptake ratio.

The two ILEs were executed on the robotic platform using an ECS-controlled workflow. After data processing and filtering of low-quality chromatographic signals (see~\Cref{sitab:TabS1} in the Supplemental information), approximately \num{1000} TMIDs per data set remained. Measurement uncertainties were consistent with the first ILE (see~\Cref{sifig:FigS6} in the Supplemental information), allowing reuse of the established STD thresholding strategy. Transient labeling profiles exhibited rich dynamics (see~\Cref{sifig:FigS4} in the Supplemental information), including pronounced overshoot behavior for several amino acids (ALA, ASP, HSER, SER), indicating high information content.

Net flux estimates obtained from the two ILEs were largely consistent with each other and with the 1‑\textsuperscript{13}C ILE, particularly for core pathways including the TCA cycle, glyoxylate shunt (GLX), gluconeogenesis (EMP), and PPP (see~\Cref{fig:figure-3}, as well as \Cref{sifig:FigS8} and \Cref{sifig:FigS9} in the Supplemental information). As expected, anaplerotic fluxes and the \texttt{PK\_PPS} flux remained unresolved due to structural non-identifiability. Notably, the analysis of the two parallel ILEs revealed pronounced differences in the inferred fraction of unlabeled ethanol uptake, differing by nearly three orders of magnitude between the tracers (\SI{0.0001 +- 0.0365}{} for the 2‑\textsuperscript{13}C ILE versus \SI{0.0808 +- 0.0201}{} for the U‑\textsuperscript{13}C ILE; see~\Cref{fig:figure-3}C). This difference likely contributes to the slightly lower EMP and PPP flux estimates for the 2‑\textsuperscript{13}C ILE. Despite these differences, overall net flux estimates were found to be robust across data sets. Overall, the evaluation of both datasets showed improved flux identifiability compared to the 1‑\textsuperscript{13}C ILE., increasing the proportion of identifiable net fluxes from \num{80}\% to \num{94}\% (U‑\textsuperscript{13}C) and \num{84}\% (2‑\textsuperscript{13}C), accompanied by a reduction of flux CI width by a factor of \num{1.3} and \num{1.2}, respectively.

In contrast to fluxes, only a minority of pool sizes could be reliably estimated: \num{25}\% and \num{27}\% were (at least weakly) identifiable for the 2- and U-\textsuperscript{13}C ILEs, respectively, compared to \num{20}\% in the 1-\textsuperscript{13}C experiment. Among the identifiable pool sizes, only a subset (4 out of 17) showed agreement within their CIs, while the majority exhibited substantial deviations, in some cases exceeding an order of magnitude (see~\Cref{fig:figure-3}B and \Cref{sifig:FigS11}, ~\cref{sitab:TabS4} in the Supplemental information). These discrepancies suggest that, unlike net fluxes, pool size estimates are sensitive to subtle experimental variations or hidden factors not captured in the model. Importantly, this behavior contrasts with multi-dataset IST \cmfa{}, where combining datasets typically improves parameter consistency. In INST \cmfa{}, joint evaluation of parallel datasets therefore requires particular care.
 
To assess whether joint evaluation of both parallel INST datasets resolves the observed discrepancies while further improving flux resolution, the \cmfa{} model was extended for multi-dataset inference. This extension accounted for dataset-specific differences in selected pool sizes (AKG, ALA, ASP, CITR, DAP, GLU, HSER, LYS, and ORN) and allowed for distinct uptake ratios of labeled versus unlabeled ethanol (see~\Cref{sifig:FigS10} in the Supplemental information). The joint analysis reduced CIs for most net fluxes (see~\Cref{fig:figure-3}), thereby confirming and strengthening the robustness of flux estimates obtained from individual ILE evaluations. In contrast, pool size estimates did not converge under joint inference. While previously inferred values were largely preserved, no additional pool size parameter became identifiable, and discrepancies in dataset-specific pool size estimates persisted. This indicates that, unlike in IST \cmfa, combining multiple INST datasets can reveal underlying variability, without necessarily reconciling all parameters.

Taken together, the multi-dataset INST \cmfa{} analysis demonstrates that iterative experimental refinement and complementary tracer design substantially improve flux inferences. At the same time, the observed variability in pool size estimates highlights a fundamental difference to IST \cmfa: while fluxes are robustly constrained by network structure and labeling dynamics, pool size inference remains sensitive to hidden effects in the absence of direct concentration measurements. This finding underscores the need for tailored strategies in multi-dataset INST \cmfa{}, both in ED and in statistical inference.

%%%%%%%%%%%%%%%%%%%%%%%%%%%%%%%%%%%%%%%%%%%%%%%%%%%%%%%%%%%%%%%%%%%%%%%%%%%%%%%%%%%%%%%%%%%%%%%%%%%%
\subsection*{Contextualization of \CG{}'s evolved metabolism on ethanol}
\label{subsec: INST-MFA discussion}
%%%%%%%%%%%%%%%%%%%%%%%%%%%%%%%%%%%%%%%%%%%%%%%%%%%%%%%%%%%%%%%%%%%%%%%%%%%%%%%%%%%%%%%%%%%%%%%%%%%%

The ethanol uptake rate in the joint evaluation was estimated with high precision ($7.69 \pm 0.33$ \SI{}{mmol.g_{X}^{-1}.h^{-1}}), thereby improving downstream flux precision across the network (see~\Cref{fig:figure-3}).
A substantial fraction of carbon entered the TCA cycle via citrate synthase (\texttt{CS}, \SI{63 +- 3}{\%}) and was partially redirected through the glyoxylate shunt (GLX, \SI{27 +- 7}{\%}), with fluxes through isocitrate lyase and malate synthase (\texttt{ICL} and \texttt{MALS} and \SI{2.16 +- 0.14}{mmol.g_{X}^{-1}.h^{-1}}, respectively). This consistently supports an active GLX during growth on ethanol. Fluxes through (ATP consuming) EMP and the PPP (\SI{13 +- 0}{\%} and \SI{20 +- 0}{\%}) remained comparatively low, but reflect stable precursor supply for biosynthesis. Notably, upper and lower metabolism appear to be primarily connected via anaplerotic reactions, as the net flux through pyruvate dehydrogenase (\texttt{PDHC}) was close to zero and directionally unresolved (see \Cref{sitab:TabS3} in the Supplemental information).

In contrast to fluxes, inferred pool sizes (see \Cref{sitab:TabS4} in the Supplemental information) showed substantially higher variability. Estimated concentrations spanned approximately four orders of magnitude, from \SI{e-4}{mmol.g_{X}^{-1}} (e.g. ASP, ORN) to \SI{e-1}{mmol.g_{X}^{-1}} (e.g. AKG). Only a subset of pool sizes (e.g. PRO, THR, GLN, GLY) was consistently estimated, whereas several metabolites exhibited large deviations, in some cases exceeding one order of magnitude (e.g. GLU, LYS). This reinforces that, unlike fluxes, pool size estimates remain sensitive to experimental and modeling uncertainties and should be interpreted with caution.

To contextualize these findings, the inferred flux map was compared to a previous IST \cmfa{} study of \CG{} WT grown on acetate~\citep{Wendisch2000}. Despite differences in model resolution and ED, the relative flux patterns showed notable agreement: the fraction of carbon entering the TCA cycle was similar (\SI{72}{\percent} on ethanol vs. \SI{76}{\percent} on acetate), as was the contribution of the glyoxylate shunt (\SI{22}{\percent} vs. \SI{18}{\percent}). 
These similarities suggest a conserved metabolic strategy for growth on C2 substrates. However, key differences arise from the higher degree of reduction of ethanol compared to acetate. The additional oxidation steps required for ethanol assimilation reduce the demand for oxidative TCA flux to (re)generate energy and reducing equivalents (NAD(P)H via \texttt{ICD} and $\alpha$-ketoglutarate dehydrogenase (\texttt{ODHC}) as well as ATP via \texttt{SCS}), leading to lower carbon loss via decarboxylation. This is consistent with a significantly higher biomass yield observed for growth on ethanol (\SI{0.45}{g_{CDW}.g_{S}^{-1}} or \SI{0.35}{g_{C_{X}}.g_{C_{S}}^{-1}} assuming \SI{0.408}{g_{C_{X}}.g_{CDW}}) compared to acetate (\SI{0.28}{g_{C_{X}}.g_{C_{S}}^{-1}})~\citep{Halle2023, Wendisch2000}. Consistent with this interpretation, gluconeogenic and PPP fluxes remained largely unchanged, reflecting stable biosynthetic precursor requirements across substrates.

Finally, the estimated GLX pool size on ethanol was markedly higher than values reported for the wild-type strain under glucose conditions (\SI{39.5 +- 21.3}{mM} vs. \SI{0.015 +- 0.014}{mM})~\citep{Tillack2012}, where the shunt is inactive, further underscoring its central role during growth on ethanol.

%%%%%%%%%%%%%%%%%%%%%%%%%%%%%%%%%%%%%%%%%%%%%%%%%%%%%%%%%%%%%%%%%%%%%%%%%%%%%%%%%%%%%%%%%%%%%%%%%%%%
\section*{DISCUSSION}
\label{sec:discussion}
%%%%%%%%%%%%%%%%%%%%%%%%%%%%%%%%%%%%%%%%%%%%%%%%%%%%%%%%%%%%%%%%%%%%%%%%%%%%%%%%%%%%%%%%%%%%%%%%%%%%

The development of a miniaturized and automated workflow for INST \cmfa{} represents a step change in how microbial metabolism can be interrogated. By reducing experimental costs and enabling parallel, automated experimentation, this approach facilitates reproducible and targeted fluxomics with scalable throughput. It thereby provides a practical route to integrate high-resolution flux analyses into iterative strain engineering and biofoundry pipelines, accelerating the knowledge-driven development of microbial production hosts for a sustainable bioeconomy.

Beyond throughput, our study provides new insights into the evaluation of multiple datasets in INST \cmfa. The presented workflow enables systematic generation and joint evaluation of parallel ILE datasets, revealing that combining complementary tracer experiments consistently improves flux precision, even in the absence of direct intracellular concentration measurements. At the same time, the analysis demonstrates that while net flux estimates remain robust across datasets, pool size inferences do not necessarily converge under joint evaluation and instead expose underlying variability. This poses a fundamental difference to IST \cmfa.

This variability likely reflects a combination of subtle experimental differences and unobserved biological factors, such as fluctuations in batch conditions or extracellular metabolite composition, which cannot be fully controlled even in automated, parallelized setups. As a consequence, pool size estimates remain sensitive to such hidden effects and should be interpreted with caution in the absence of direct measurements.

Looking ahead, several challenges remain for scalable INST \cmfa. Multi-dataset inference does not yet fully account for all sources of bias across experiments, highlighting the need for approaches that leverage parallelization not only for throughput but also for systematic bias detection and correction. In addition, current treatments of measurement uncertainty may misrepresent confidence levels and thereby affect parameter estimation. Advances in statistical inference, including Bayesian approaches~\citep{Theorell2024,Niesser2024}, offer promising avenues to address these limitations, albeit with increased demands on data, e.g. intracellular pool size measurements, and computational resources.

Finally, the presented workflow is not restricted to microtiter plate formats but can be adapted to other low-cost, automatable cultivation platforms such as Pioreactors (\url{https://pioreactor.com/}) or eVOLVER~\citep{Wong2018}, broadening access to high-throughput INST \cmfa. Together, this work establishes automated multi-dataset INST \cmfa{} as a scalable framework for quantitative flux analysis and highlights both its potential and its current limitations for future developments.

\subsection*{Data and code availability}

Specific biomass and substrate concentration measurements, raw TMIDs derived from LC-QToF-MS, Omix network models, FluxML files and original code for data evaluation are available from the GitHub repository \href{https://github.com/JuBiotech/Supplement-to-Niesser-and-Stratmann-et-al.-Trends-Biotechnol.-2026}{https://github.com/JuBiotech/Supplement-to-Niesser-and-Stratmann-et-al.-Trends-Biotechnol.-2026}.

\section*{Acknowledgements}

The work was performed as part of the Helmholtz School for Data Science in Life, Earth and Energy  (HDS-LEE) and received  funding  from  the  Helmholtz  Association  of  German Research Centres. We gratefully acknowledge support and funding from the Deutsche Forschungsgemeinschaft (project no. 560539272). We acknowledge the computing time on the supercomputer JURECA at Forschungszentrum Jülich (grant no. \texttt{hpcmfa}).

\section*{Declaration of interests}

The authors have no conflict of interest to declare.

\section*{Declaration of generative AI and AI-assisted technologies}

During the preparation of this work the authors used DeepL and ChatGPT in order to improve the readability and language of the manuscript. After using these tools, the authors carefully reviewed and edited the content as needed and take full responsibility for the content of the published article.

% \bibliography{references}
%\bibliographystyle{abbrvnat}

\bibliographystyle{abbrvnat}

%%%%%%%%%%%%%%%%%%%%%%%%%%%%%%%%%%%%%%%%%%%%%%
%%%%%%%%%%%%%%%%%%%%%%%%%%%%%%%%%%%%%%%%%%%%%%

\newpage
\appendix
\onecolumn

{
    \centering
    \textbf{Supplemental Information for:}\\[3ex]
    \textbf{\Large 
        Automated multi-dataset INST \textsuperscript{13}C metabolic flux analysis at microliter scale reveals robust fluxes but variable metabolite pools in \textit{Corynebacterium glutamicum}
        \vspace*{\baselineskip}\\[5ex]
    }
 }

\maketitle

\section*{\normalsize List of Supplemental Figures}
\footnotesize
\noindent
\textbf{Figure S1.} Reproduction of the hot isopropanol quenching validation experiment reported by \citet{Niesser2022} using six biological replicates.

\vspace{0.5em}

\noindent
\textbf{Figure S2.} Business Process Model and Notation (BPMN) 2.0-inspired flow scheme of the automated isotope labeling experiment (ILE) workflow.

\vspace{0.5em}

\noindent
\textbf{Figure S3.} Interpreting transient labeling dynamics in the context of the metabolic network.

\vspace{0.5em}

\noindent
\textbf{Figure S4.} Selected labeling incorporation dynamics for the 1-, 2-, and U-\textsuperscript{13}C ethanol ILE datasets after natural isotope abundance correction.

\vspace{0.5em}

\noindent
\textbf{Figure S5.} Distribution of standard deviations (STDs) of TMID fractional enrichments for the three INST-ILE datasets.

\vspace{0.5em}

\noindent
\textbf{Figure S6.} Relationship between TMID fractional enrichments and their corresponding standard deviations (STDs) for the three INST-ILE datasets.

\vspace{0.5em}

\noindent
\textbf{Figure S7.} Flux map of \CG{} WT\_EtOH-Evo obtained from the evaluation of the 1-\textsuperscript{13}C ethanol INST-ILE dataset.

\vspace{0.5em}

\noindent
\textbf{Figure S8.} Flux map of \CG{} WT\_EtOH-Evo obtained from the individual evaluation of the 2-\textsuperscript{13}C ethanol INST-ILE dataset.

\vspace{0.5em}

\noindent
\textbf{Figure S9.} Flux map of \CG{} WT\_EtOH-Evo obtained from the individual evaluation of the U-\textsuperscript{13}C ethanol INST-ILE dataset.

\vspace{0.5em}

\noindent
\textbf{Figure S10.} Flux map of \CG{} WT\_EtOH-Evo obtained from the joint evaluation of the 2- and U-\textsuperscript{13}C ethanol INST-ILE datasets.

\vspace{0.5em}

\noindent
\textbf{Figure S11.} Comparison of identifiable pool size estimates obtained from independent evaluations of the 1-, 2-, and U-\textsuperscript{13}C INST-ILE datasets, as well as from the joint evaluation of the 2- and U-\textsuperscript{13}C datasets.

\vspace{2em}

\section*{\normalsize List of Supplemental Tables}

\textbf{Table S1.} Overview of the filtered transient labeling datasets used for INST \cmfa{}.

\vspace{0.5em}

\noindent
\textbf{Table S2.} Model parameter statistics.

\vspace{0.5em}

\noindent
\textbf{Table S3.} Absolute net flux estimates obtained from independent analyses of the 1-, 2-, and U-\textsuperscript{13}C INST-ILE datasets, as well as from the joint evaluation of the 2- and U-\textsuperscript{13}C datasets.

\vspace{0.5em}

\noindent
\textbf{Table S4.} Pool size estimates obtained from independent analyses of the 1-, 2-, and U-\textsuperscript{13}C INST-ILE datasets, as well as from the joint evaluation of the 2- and U-\textsuperscript{13}C datasets.

\normalsize
\clearpage

\mbox{}

%%%%%%%%%%%%%%%%%%%%%%%%%%%%%%%%%%%%%%%%
%%%%%   FIGURES
%%%%%%%%%%%%%%%%%%%%%%%%%%%%%%%%%%%%%%%%

\begin{figure}[htp!]
    \centering
    \includegraphics[width=0.65\textwidth]{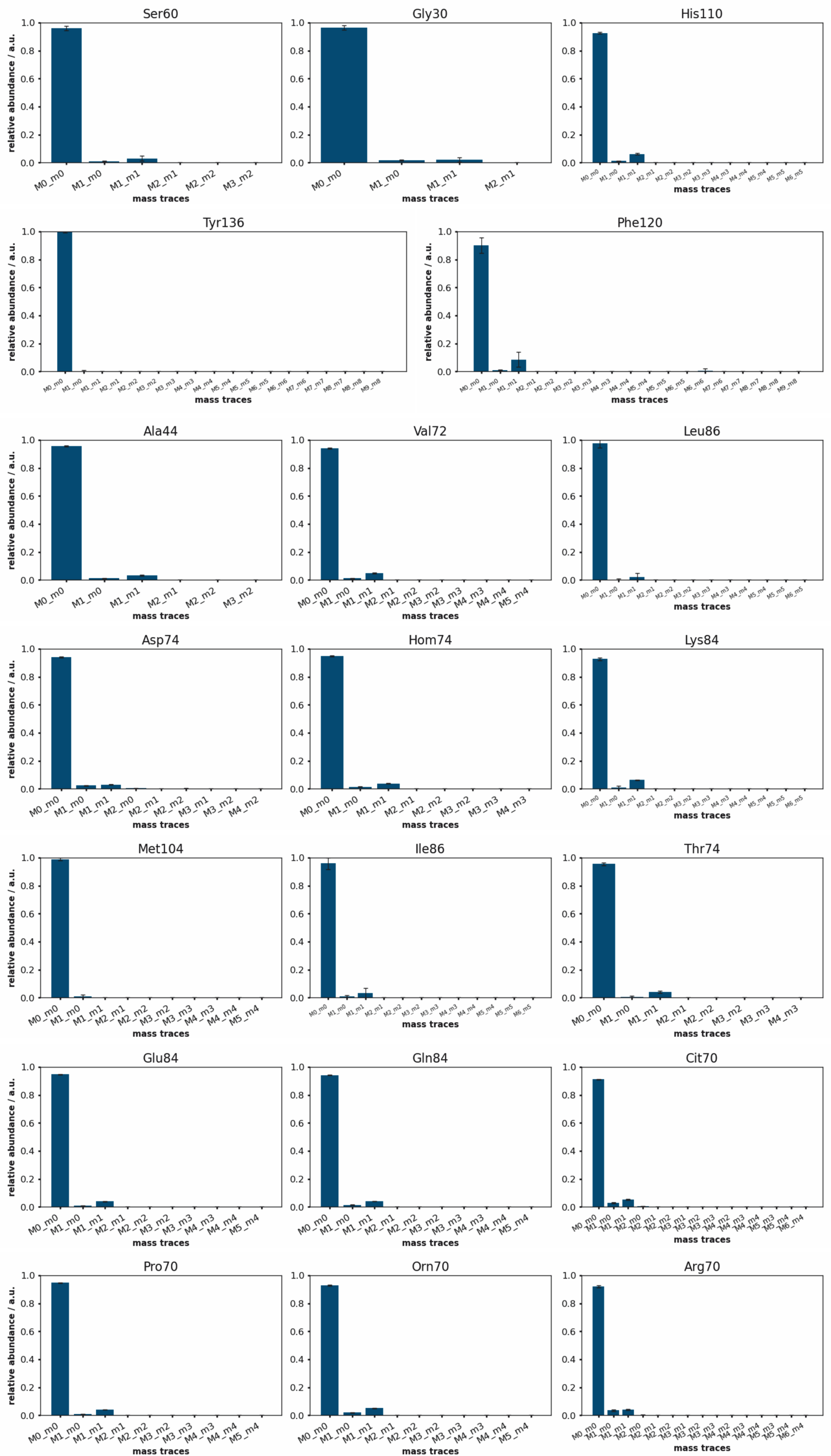}
    \caption{\textbf{Reproduction of the hot isopropanol quenching validation experiment reported by \citet{Niesser2022} using six biological replicates.} U-\textsuperscript{13}C glucose was added directly to the quenching reagent instead of the cultivation medium, such that any label incorporation into amino acids beyond natural labeling would indicate residual enzyme activity during quenching. As no measurable fractional enrichment was detected, the systematic bias discussed in the original study could be excluded.}
    \label{sifig:FigS1}
\end{figure}

\clearpage
\FloatBarrier

\begin{figure}[htp!]
    \centering
    \includegraphics[width=0.95\textwidth]{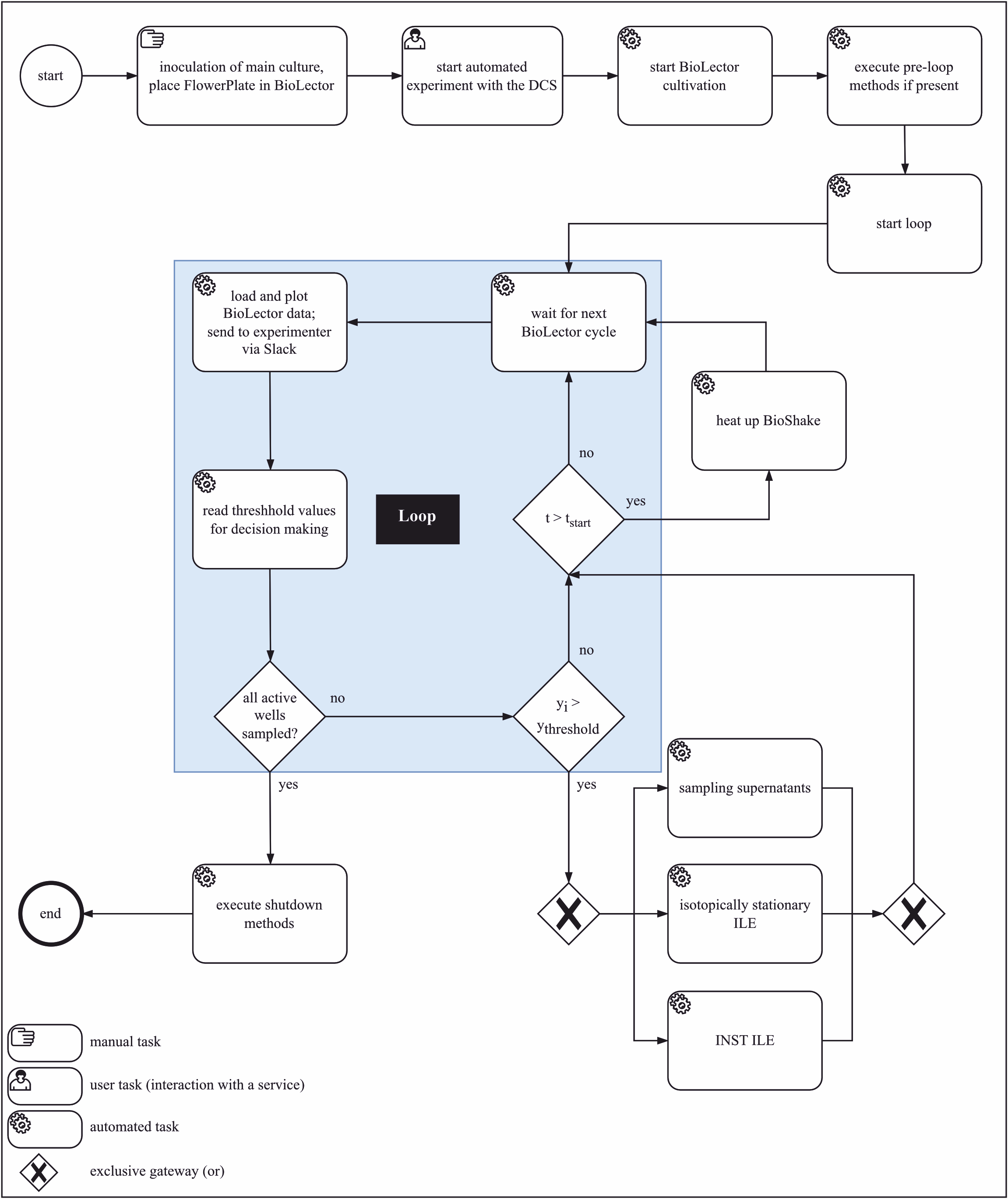}
    \caption{\textbf{Business Process Model and Notation (BPMN) 2.0-inspired flow scheme of the automated isotope labeling experiment (ILE) workflow.} During the pre-experimental phase, a Python-based experimental control script (ECS) is generated and pipetting schemes are defined according to the experimental design, including the number and placement of sampling time points and conditions on the microtiter plate.}
    \label{sifig:FigS2}
\end{figure}

\clearpage
\FloatBarrier

\begin{figure}[htp!]
    \centering
    \includegraphics[width=0.82\linewidth]{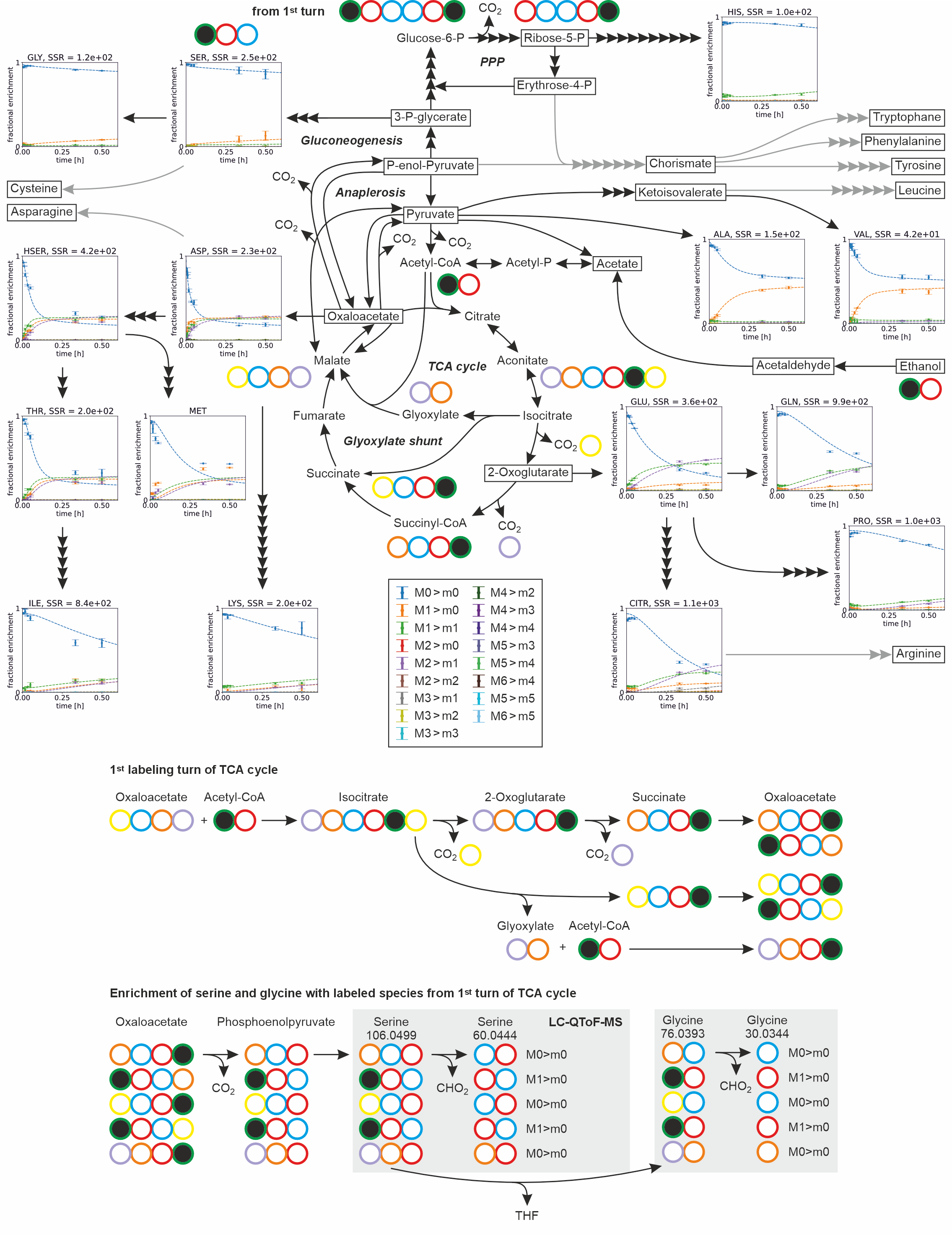}
    \caption{\textbf{Interpreting transient labeling dynamics in the context of the metabolic network.} Upon introducing 1-\textsuperscript{13}C ethanol, glyoxylate shunt activity and label scrambling during succinate-to-fumarate interconversion generate five labeled oxaloacetate isotopomers. During gluconeogenesis, phosphoenolpyruvate carboxykinase-mediated decarboxylation produces phosphoenolpyruvate- and serine-derived isotopomers labeled at the first carbon atom position. In LC-QToF-MS analysis, fragmentation of the intact serine [M+H]\textsuperscript{+} ion ($m/z=106$) yields an unlabeled product ion. for 1-\textsuperscript{13}C ethanol, only the serine $M1>m0$ transition increases gradually over time, independent of the number of turns through the TCA cycle. 
    In contrast, the PPP-derived amino acid histidine showed only marginal label incorporation during the observed time window. This is consistent with the long biosynthetic distance from ribose-5-phosphate, consisting of ten reaction steps prior to histidine formation. Notably, the absense of substantial labeling does not arise from decarboxylation of 6-phosphogluconate to ribose-5-phosphate within the PPP, since the sixth carbon atom of any C6 intermediate upstream of fructose-6-phosphate is retained even during cyclic PPP operation.}
    \label{sifig:FigS3}
\end{figure}

\clearpage
\FloatBarrier

\begin{figure}[htp!]
    \centering
    \includegraphics[width=0.9\linewidth]{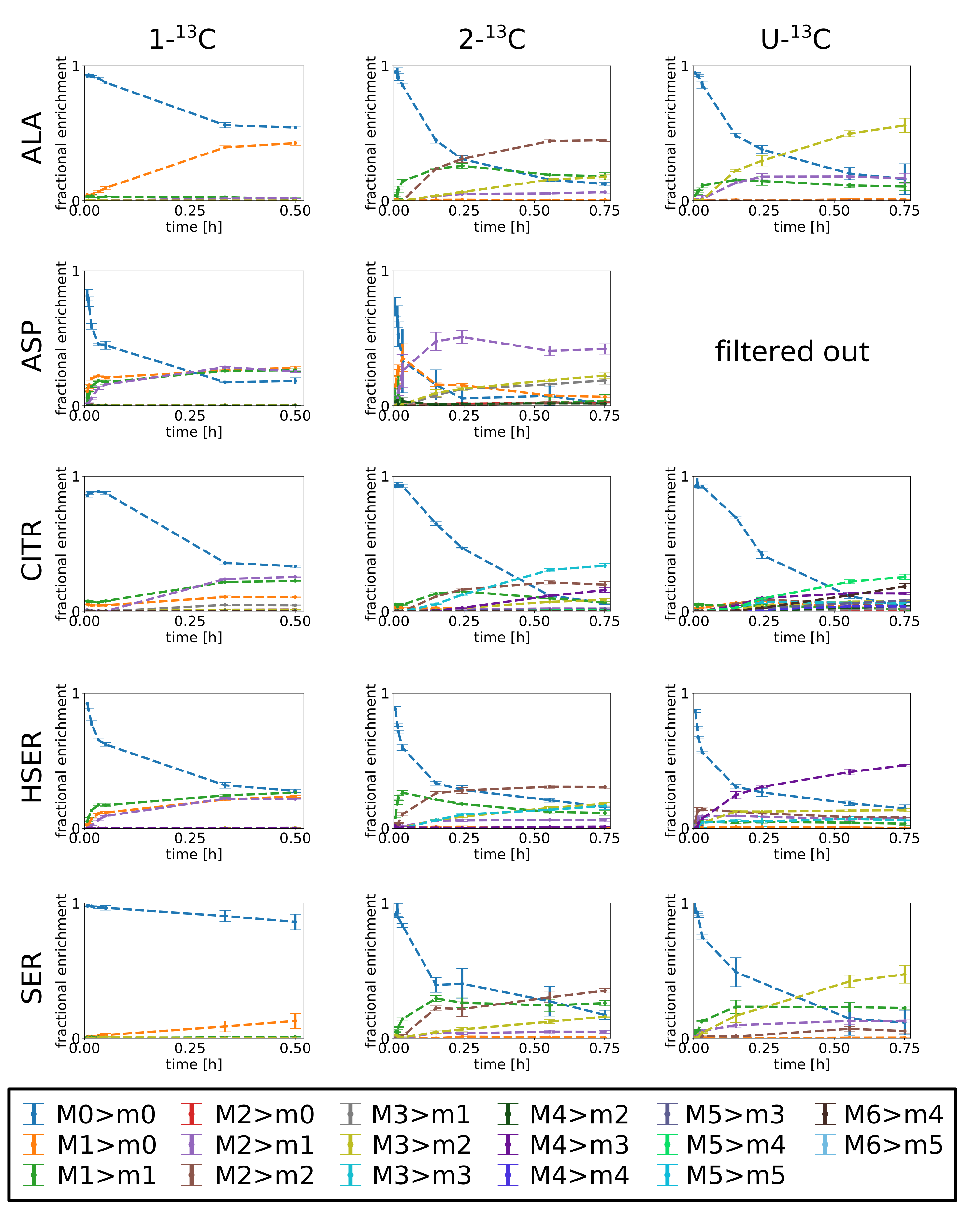}
    \caption{\textbf{Selected labeling incorporation dynamics for the 1-, 2-, and U-\textsuperscript{13}C ethanol ILE datasets after natural isotope abundance correction.} Shown are amino acids exhibiting pronounced transient profiles, including overshoot behavior. Data points represent mean values across three biological replicates, while dotted lines are included to guide the eye. Since TMIDs are the product of tandem MS, their denomination features both the m/z value ``M'' of the precursor ion in MS stage 1 and the m/z value ``m'' of the product ion in MS stage 2 followed by the number of incorporated labels and commonly separated by ``$>$''. In case of small molecules such as amino acids, the m/z value and the mass are effectively identical, as they are only protonated once. Accordingly, each label increases the m/z value by 1.}
    \label{sifig:FigS4}
\end{figure}

\clearpage
\FloatBarrier

\begin{figure}[!htp]
    \centering
    \includegraphics[width=0.75\linewidth]{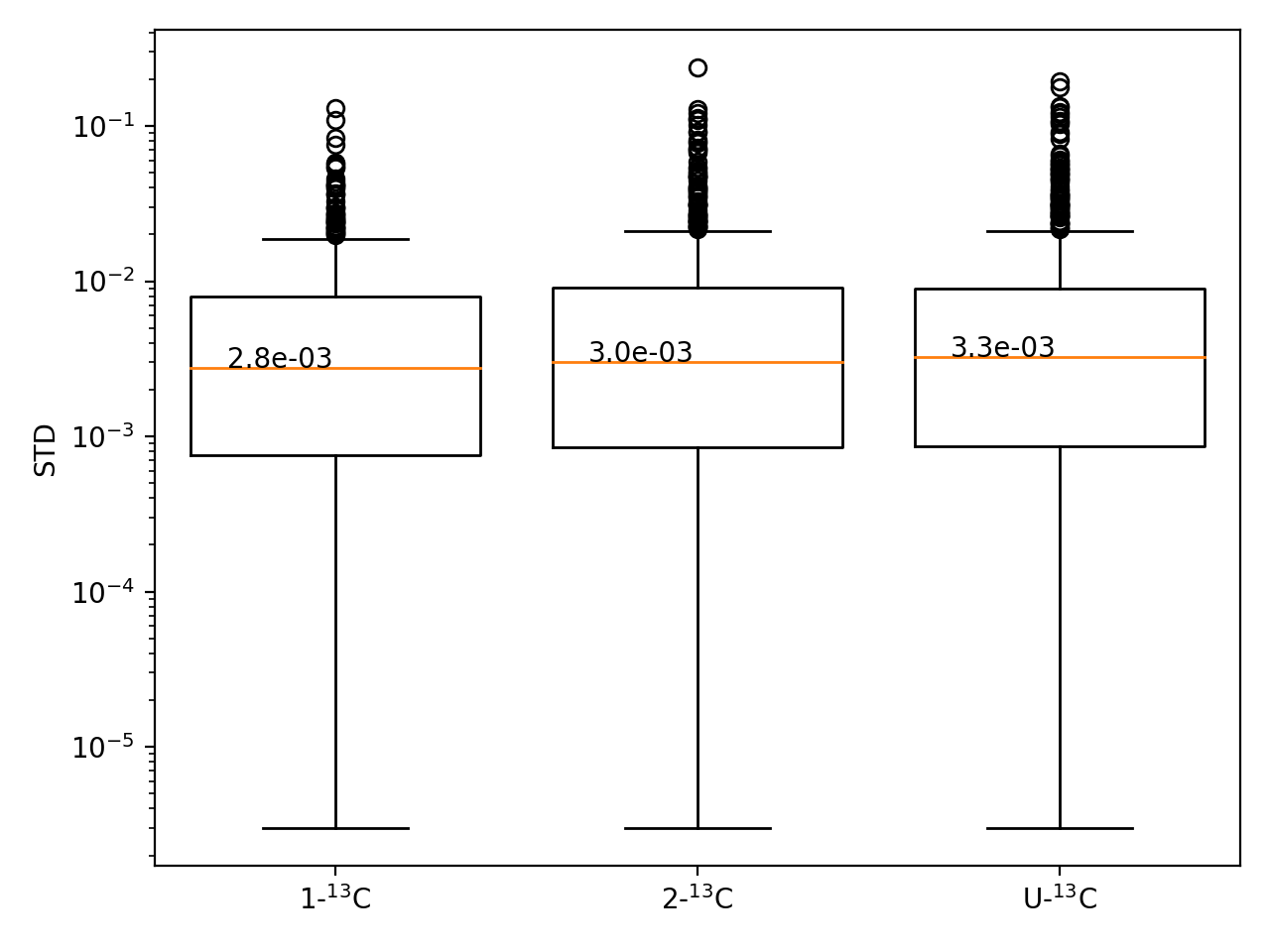}
    \caption{\textbf{Distribution of standard deviations (STDs) of TMID fractional enrichments for the three INST-ILE datasets.} All ILEs were performed in biological triplicates on the robotic platform. After removal of low-quality measurements and correction for natural isotope abundance, TMIDs were derived for the remaining amino acid fragments and STDs were calculated from the triplicate-derived fractional enrichments. The resulting STD distributions showed similar characteristics across all three datasets, with no statistically significant differences (Mann–Whitney U-test, $p > 0.05$). Approximately half of all STDs fall within the range of $10^{-2}$ to $10^{-3}$ (box). Of the remaining values, one quarter lies between $10^{-3}$ and $5 \cdot 10^{-2}$ (upper whisker), whereas another quarter is up to two orders of magnitude smaller (lower whisker). Extreme values range from $2 \cdot 10^{-1}$ down to $3 \cdot 10^{-6}$, approaching the LC-QToF-MS detection limit.}
    \label{sifig:FigS5}
\end{figure}

\clearpage
\FloatBarrier

\begin{figure}[!htp]
    \centering
    \includegraphics[width=0.75\linewidth]{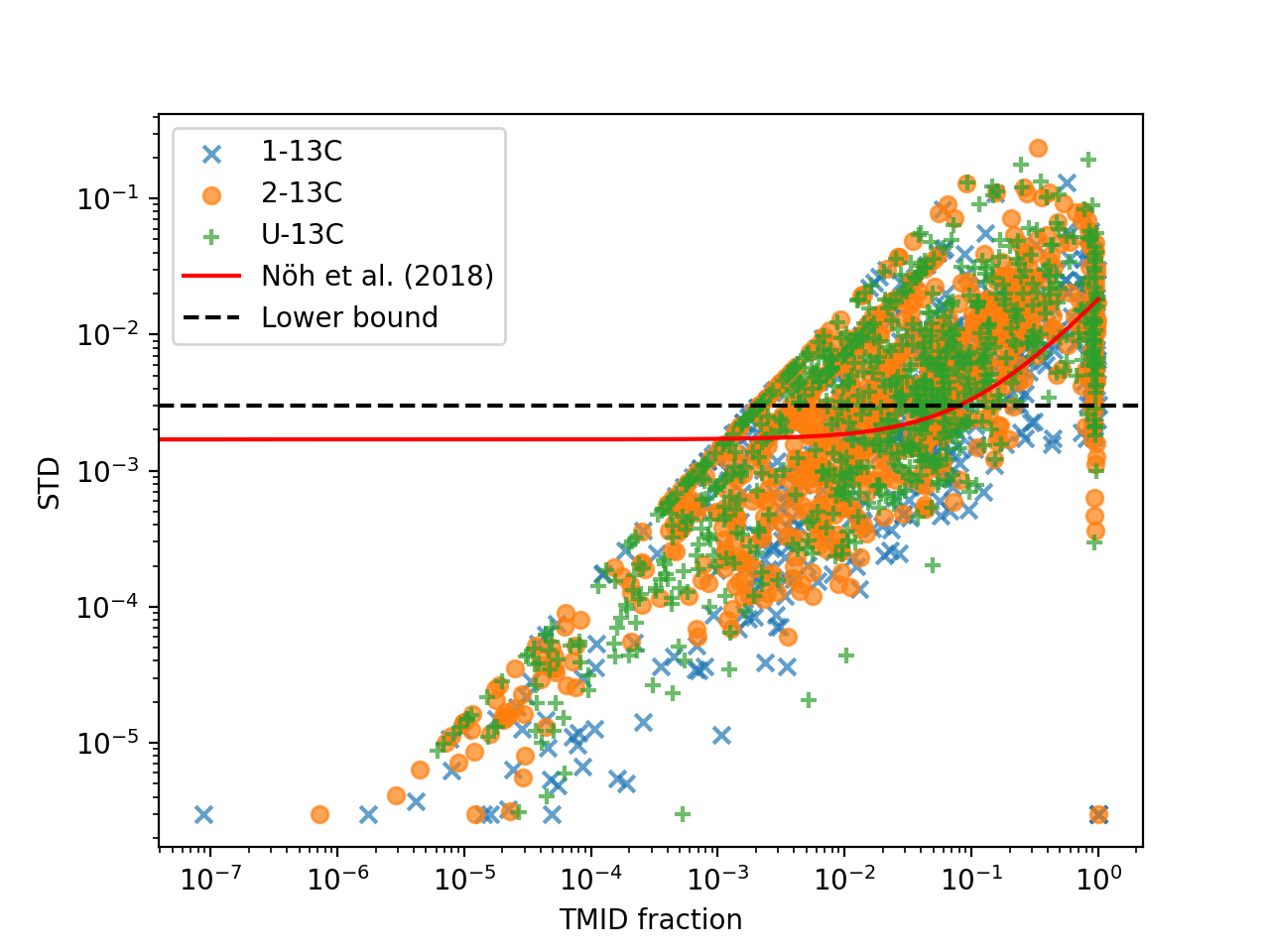}
    \caption{\textbf{Relationship between TMID fractional enrichments and their corresponding standard deviations (STDs) for the three INST-ILE datasets.} STDs generally increased with enrichment magnitude, consistent with previous observations~\citep{Yuan2010}, although several outliers occurred at high fractional enrichments close to 1.
    No systematic tracer-dependent differences in enrichment distributions were observed. Due to analytical detection limits, STDs were effectively bounded below by approximately $10^{-6}$. The red line indicates the empirical error model reported by \citet{Noh2018}, derived from a survey of LC-MS/MS platforms, metabolite classes, and analytical workflows.
    Notably, many experimentally derived STDs fell below this empirical expectation. To avoid unrealistically small uncertainties dominating parameter estimation, a lower STD threshold of $3 \cdot 10^{-3}$ was applied (black dotted line), corresponding to the median of the observed STD distributions.}
    \label{sifig:FigS6}
\end{figure}

\clearpage
\FloatBarrier

\begin{figure}[htp!]
    \centering
    \includegraphics[width=1\textwidth]{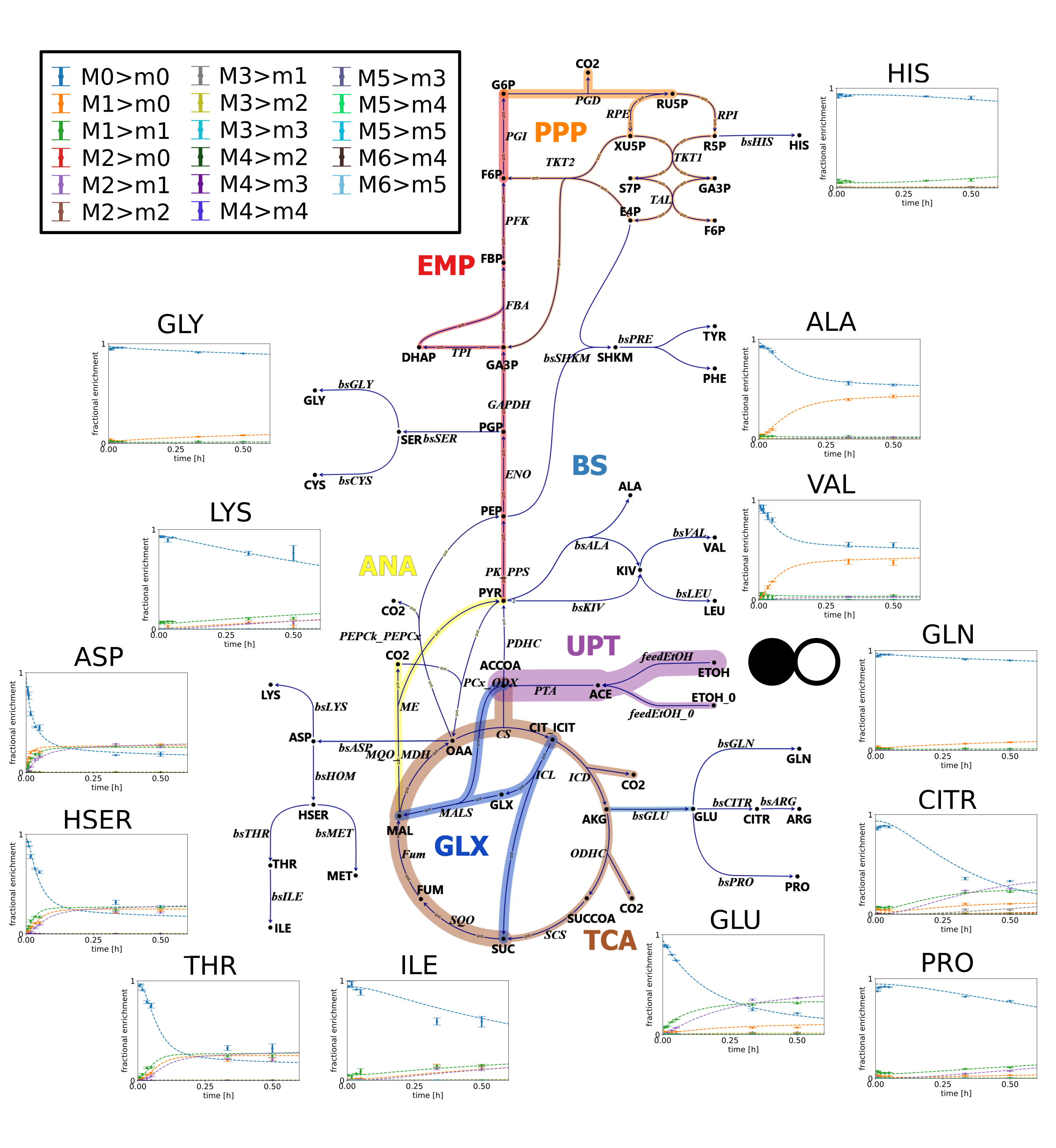}
    \caption{\textbf{Flux map of \CGcap{} WT\_EtOH-Evo obtained from the evaluation of the 1-\textsuperscript{13}C ethanol INST-ILE dataset.} Data points represent transient TMID fractional enrichments of the filtered free amino acids used for parameter inference, with whiskers indicating the corresponding thresholded STDs (see \Cref{sitab:TabS1}). Solid lines denote simulated labeling trajectories obtained from the best-fit parameter set (see~\Cref{sitab:TabS3}, \Cref{sitab:TabS4}).}
    \label{sifig:FigS7}
\end{figure}

\clearpage
\FloatBarrier

\begin{figure}[htp!]
    \centering
    \includegraphics[width=1\textwidth]{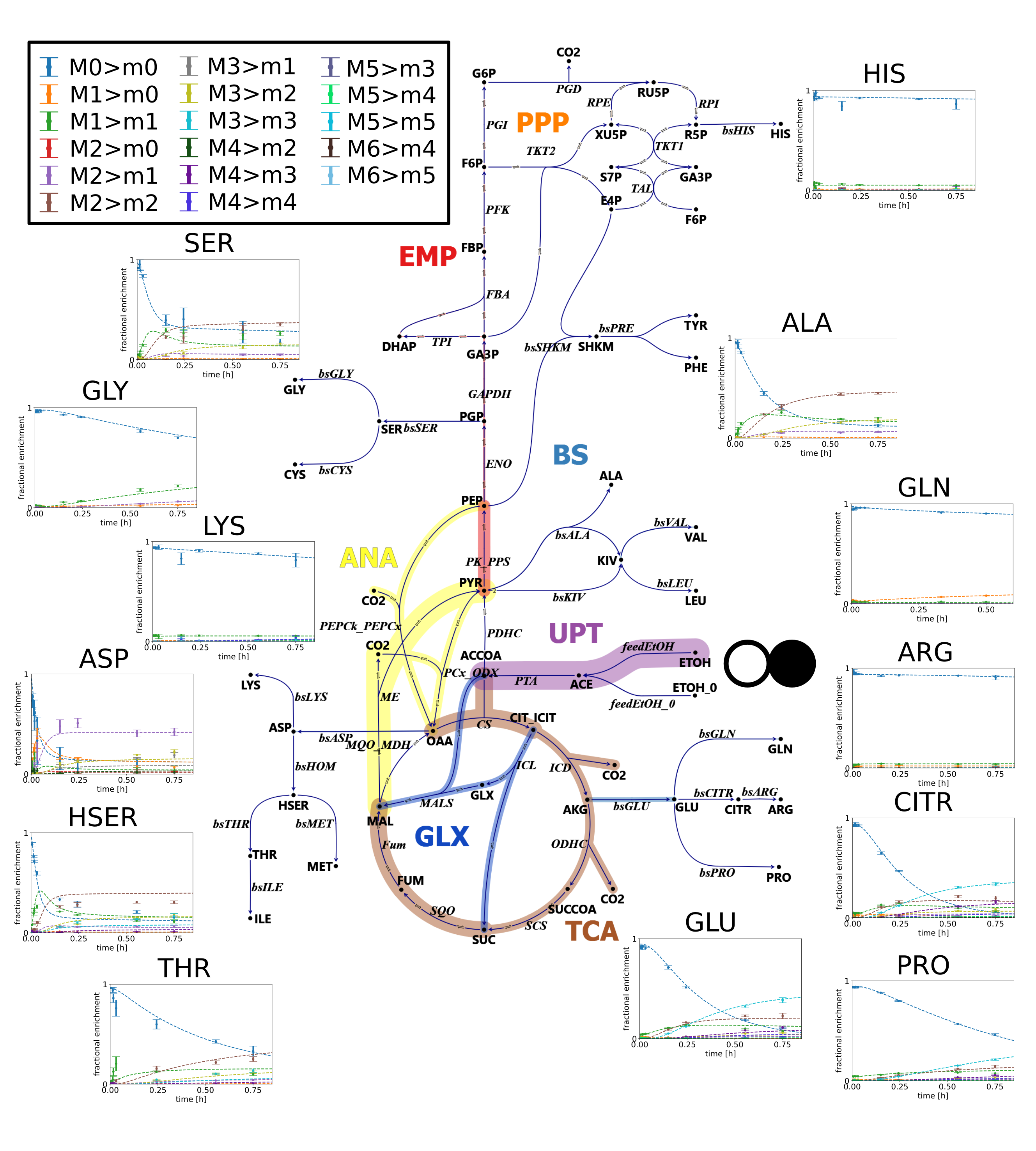}
    \caption{\textbf{Flux map of \CGcap{} WT\_EtOH-Evo obtained from the individual evaluation of the 2-\textsuperscript{13}C ethanol INST-ILE dataset.} Data points represent transient TMID fractional enrichments of the filtered free amino acids (see \Cref{sitab:TabS1}) used for parameter inference, with whiskers indicating the corresponding thresholded STDs. Solid lines denote simulated labeling trajectories obtained from the best-fit parameter set (see~\Cref{sitab:TabS3}, \Cref{sitab:TabS4}).}
    \label{sifig:FigS8}
\end{figure}

\clearpage
\FloatBarrier

\begin{figure}[htp!]
    \centering
    \includegraphics[width=1\textwidth]{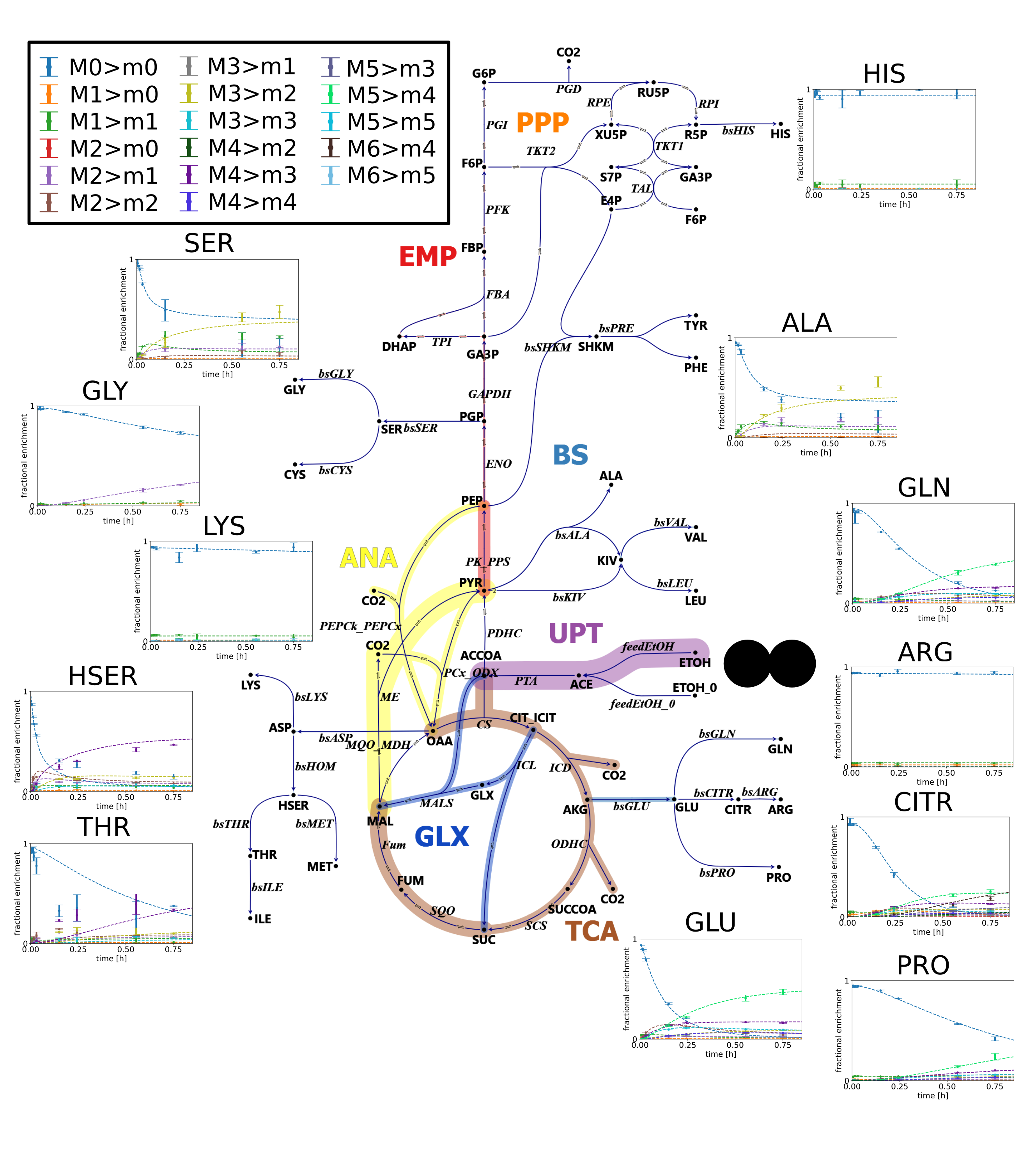}
    \caption{\textbf{Flux map of \CGcap{}{} WT\_EtOH-Evo obtained from the individual evaluation of the U-\textsuperscript{13}C ethanol INST-ILE dataset.} Data points represent transient TMID fractional enrichments of the filtered free amino acids (see \Cref{sitab:TabS1}) used for parameter inference, with whiskers indicating the corresponding thresholded STDs. Solid lines denote simulated labeling trajectories obtained from the best-fit parameter set (see~\Cref{sitab:TabS3}, \Cref{sitab:TabS4}).}
    \label{sifig:FigS9}
\end{figure}

\clearpage
\FloatBarrier

\begin{figure}[htp!]
    \centering
    \includegraphics[width=0.75\textwidth]{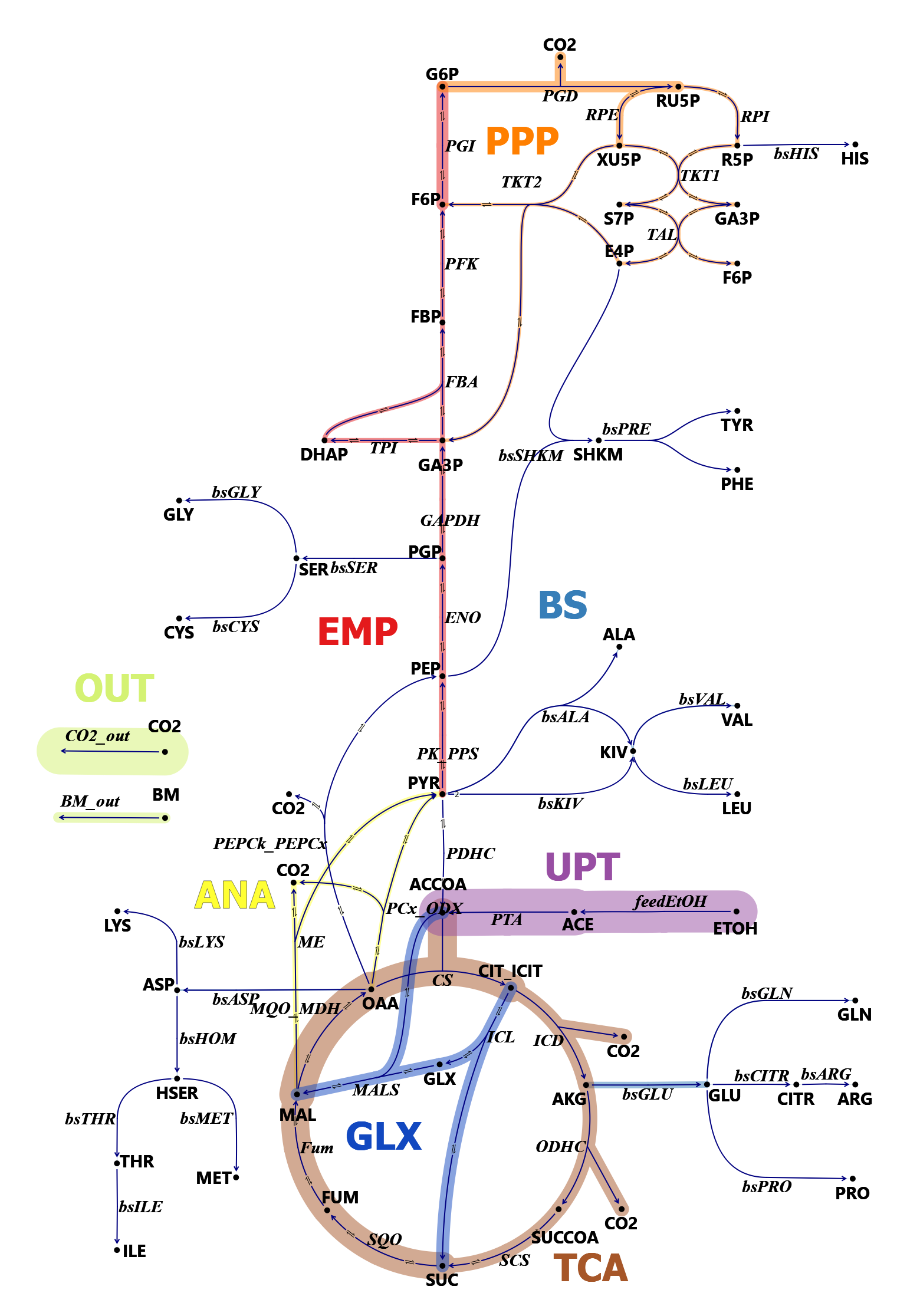}
    \caption{\textbf{Flux map of \CGcap{} WT\_EtOH-Evo obtained from the joint evaluation of the 2- and U-\textsuperscript{13}C ethanol INST-ILE datasets.} For multi-dataset inference, the corresponding single-ILE \cmfa{} models were programmatically merged into a combined model comprising two tracer-specific sub-models. Net flux parameters were constrained to identical values across both sub-models because their estimates were consistent in the independent evaluations, whereas selected metabolite pool sizes (AKG, ALA, ASP, CITR, CO2, DAP, DHAP, E4P, F6P, FBP, G6P, GA3P, GLU, HIS, HSER, LYS, ORN, PGP, PRA, R5P, RU5P, VAL, and XU5P) were retained as ILE-specific parameters due to divergent single-dataset estimates (see~\Cref{sifig:FigS11}). Reaction reversibilities and uptake ratios of labeled versus unlabeled ethanol were modeled independently for each ILE. The best-fit parameter set is given in~\Cref{sitab:TabS3} and \Cref{sitab:TabS4}.}
    \label{sifig:FigS10}
\end{figure}

\clearpage
\FloatBarrier

\begin{figure}[!htp]
    \centering
    \includegraphics[width=0.6\linewidth]{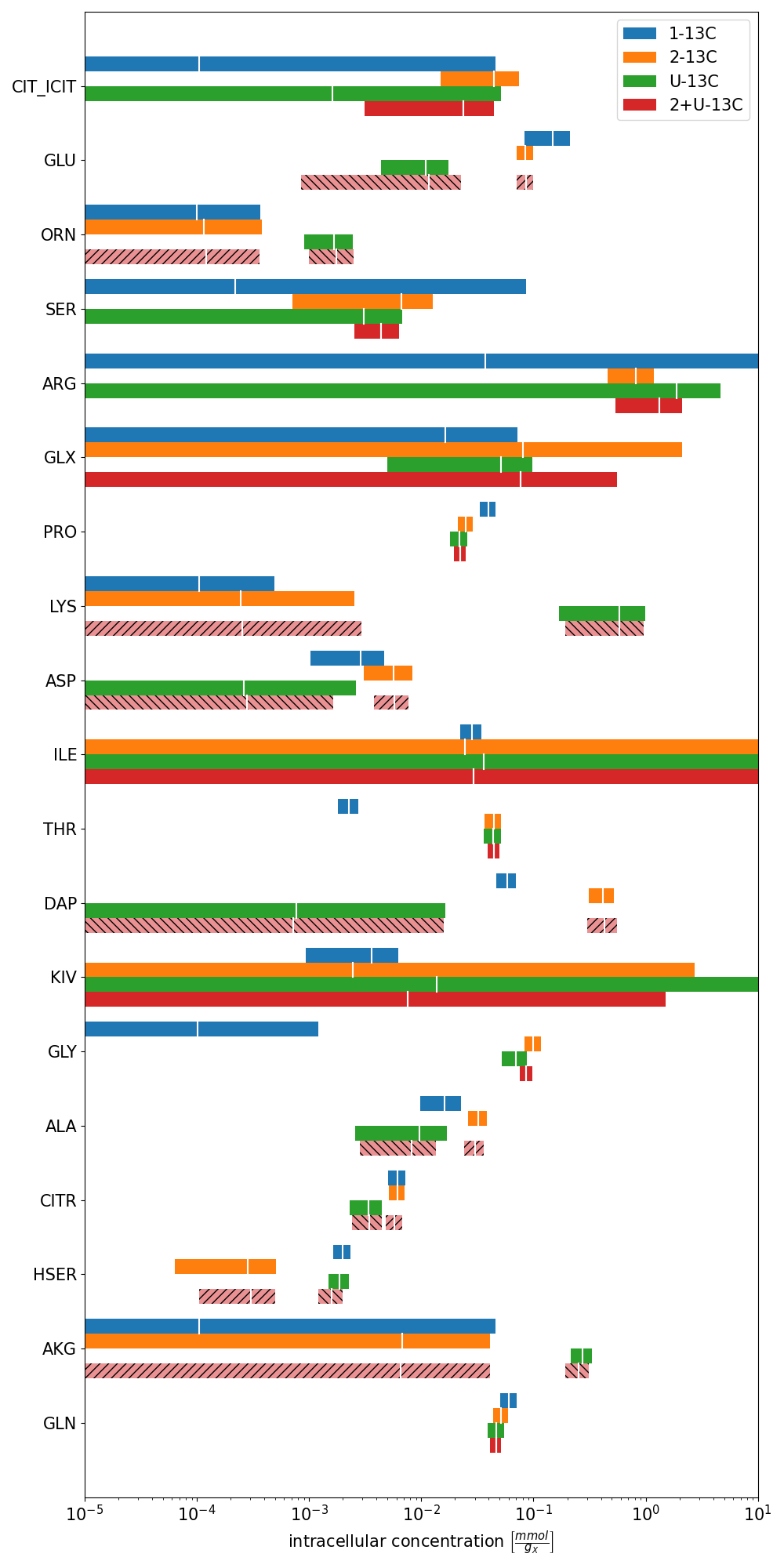}
    \caption{\textbf{Comparison of identifiable pool size estimates obtained from independent evaluations of the 1-, 2-, and U-\textsuperscript{13}C INST-ILE datasets, as well as from the joint evaluation of the 2- and U-\textsuperscript{13}C datasets.} Pool sizes are grouped according to the agreement between dataset-specific estimates: (1) all datasets agree (GLN); (2) the 1- and 2-\textsuperscript{13}C datasets agree, whereas the U-\textsuperscript{13}C dataset deviates (AKG, ASP, CITR, GLU, LYS, ORN); (3) the 1- and U-\textsuperscript{13}C datasets agree, whereas the 2-\textsuperscript{13}C dataset deviates (ALA, HSER); (4) the 2- and U-\textsuperscript{13}C datasets agree, whereas the 1-\textsuperscript{13}C dataset deviates (GLY, THR, PRO); and (5) all datasets differ or remain inconclusive (DAP, KIV, ILE, GLX, ARG, SER, CIT\_ICIT).}
    \label{sifig:FigS11}
\end{figure}

\clearpage
\FloatBarrier

%%%%%%%%%%%%%%%%%%%%%%%%%%%%%%%%%%%%%%%%
%%%%%   TABLES
%%%%%%%%%%%%%%%%%%%%%%%%%%%%%%%%%%%%%%%%

\begin{table}[htp!]
    \caption{\textbf{Overview of the filtered transient labeling datasets used for INST \cmfa{}.} For each  ILE, amino acid fragments were manually inspected and measurements with consistently insufficient signal quality were excluded from further analysis. This affected ASN, CYS, LEU, PHE, TRP, and TYR across all datasets. Additional measurements were removed when they exhibited irregular or discontinuous labeling trajectories incompatible with biologically plausible enrichment dynamics. Specifically, ARG and ORN were excluded from the 1-\textsuperscript{13}C dataset, ORN and VAL from the 2-\textsuperscript{13}C dataset, and ASP, ORN, and VAL from the U-\textsuperscript{13}C dataset. For the 1-\textsuperscript{13}C ILE, MET labeling trajectories additionally showed persistent disagreement with model-based simulations and were therefore excluded. STD thresholding is described in \Cref{sifig:FigS6}. Total labeling measurements comprise all filtered TMID observations including biological replicates. Independent measurements correspond to retained observations after condensing triplicate TMIDs into mean values. Non-redundant measurements denote the remaining unique TMID entries after removal of one redundant component per time point and amino acid measurement group arising from the fractional nature of the distributions. Each dataset additionally includes two extracellular rate measurements: the specific growth rate ($\mu$) and the specific ethanol uptake rate ($q_{\mathrm{EtOH}}$).}
    \centering
    \begin{tabular}{lcccc}
        \toprule
        ILE &
        time points &
        \makecell{total \# \\ labeling \\ measurements} &
        \makecell{\# retained \\ labeling \\ measurements} &
        \makecell{\# non-redundant \\ labeling \\ measurements} \\
        \midrule
        
        1-\textsuperscript{13}C &
        \makecell{
        \SI{24}{s}, \SI{35}{s}, \SI{60}{s}, \SI{120}{s}, \\
        \SI{180}{s}, \SI{1200}{s}, \SI{1800}{s}
        } &
        \SI{2772}{} &
        \SI{924}{} &
        \SI{826}{} \\
        
        2-\textsuperscript{13}C &
        \makecell{
        \SI{20.6}{s}, \SI{47.9}{s}, \SI{59.4}{s}, \\
        \SI{111.3}{s}, \SI{540.6}{s}, \SI{875.6}{s}, \\
        \SI{1995.8}{s}, \SI{2703.2}{s}
        } &
        \SI{3036}{} &
        \SI{1016}{} &
        \SI{904}{} \\
        
        U-\textsuperscript{13}C &
        ditto &
        \SI{2727}{} &
        \SI{922}{} &
        \SI{826}{} \\

        2- \& U-\textsuperscript{13}C &
        ditto & 
        \SI{5763}{} &
        \SI{1938}{} &
        \SI{1912}{} \\
        \bottomrule
    \end{tabular}
    \label{sitab:TabS1}
\end{table}

\clearpage
\FloatBarrier

\begin{table}[htp!]
    \caption{\textbf{Model parameter statistics.} Numbers denote the numbers of independent parameters of the respective \cmfa{} models, while values in brackets indicate the total number of parameters contained in each model formulation.}
    \centering
    \begin{tabular}{lcccc}
        \toprule
        Configuration &
        \# net fluxes & 
        \# xch fluxes &
        \# pool sizes &
        $\sum$ \\
        \midrule
        
        1-\textsuperscript{13}C &
        \SI{7}{} (\SI{107}{}) &
        \SI{22}{} (\SI{22}{}) &
        \SI{56}{} (\SI{56}{}) &
        \SI{85}{} (\SI{185}{}) \\
        
        2-\textsuperscript{13}C &
        \SI{7}{} (\SI{107}{}) &
        \SI{22}{} (\SI{22}{}) &
        \SI{56}{} (\SI{56}{}) &
        \SI{85}{} (\SI{185}{}) \\
        
        U-\textsuperscript{13}C &
        \SI{7}{} (\SI{107}{}) &
        \SI{22}{} (\SI{22}{}) &
        \SI{56}{} (\SI{56}{}) &
        \SI{85}{} (\SI{185}{}) \\

        2- \& U-\textsuperscript{13}C &
        \SI{8}{} (\SI{108}{}) &
        \SI{44}{} (\SI{44}{}) &
        \SI{79}{} (\SI{79}{}) &
        \SI{131}{} (\SI{231}{}) \\
        \bottomrule
    \end{tabular}
    \label{sitab:TabS2}
\end{table}

\clearpage
\FloatBarrier

\begin{landscape}
\begin{longtable}{lcccc}
    
        \caption{\textbf{Absolute net flux estimates obtained from independent analyses of the 1-, 2-, and U-\textsuperscript{13}C INST-ILE datasets, as well as from the joint evaluation of the 2- and U-\textsuperscript{13}C datasets.} Fluxes are given in \SI{}{mmol.g_{X}^{-1}.h^{-1}}. For each ILE, the reported values correspond to the best-fit estimate followed by the associated 95\% CI in brackets. During the estimation, the absolute values of net fluxes were constrained to be less than twice the ethanol uptake rate. The corresponding flux maps are provided in \Cref{sifig:FigS7}, \Cref{sifig:FigS8}, \Cref{sifig:FigS9}, 
        and \Cref{sifig:FigS10}.} 
        \\
        
        \toprule
        \textbf{Net flux} & 
        \textbf{1-\textsuperscript{13}C} & 
        \textbf{2-\textsuperscript{13}C} & 
        \textbf{U-\textsuperscript{13}C} & 
        \textbf{2- \& U-\textsuperscript{13}C} \\
        \midrule
        \endfirsthead

        \toprule
        \textbf{Net flux} & 
        \textbf{1-\textsuperscript{13}C} & 
        \textbf{2-\textsuperscript{13}C} & 
        \textbf{U-\textsuperscript{13}C} & 
        \textbf{2- \& U-\textsuperscript{13}C} \\
        \midrule
        \endhead

        \bottomrule
        \endlastfoot

        \textbf{CS} & 4.65 (0.00, 6.31) & 4.55 (3.29, 5.80) & 5.24 (4.24, 6.24) & 4.87 (4.44, 5.29) \\ 
        \textbf{ENO} & 1.41 (0.52, 6.83) & 0.62 (0.52, 1.64) & 1.29 (0.60, 1.97) & 1.02 (0.73, 1.31) \\ 
        \textbf{FBA} & 0.99 (0.16, 6.47) & 0.20 (0.16, 1.23) & 0.92 (0.21, 1.63) & 0.67 (0.35, 0.98) \\ 
        \textbf{Fum} & 4.46 (-0.19, 6.15) & 4.36 (3.10, 5.62) & 5.08 (4.07, 6.08) & 4.71 (4.28, 5.14) \\ 
        \textbf{GAPDH} & 1.24 (0.38, 6.68) & 0.45 (0.38, 1.47) & 1.14 (0.45, 1.83) & 0.88 (0.58, 1.18) \\ 
        \textbf{ICD} & 2.08 (0.22, 4.14) & 2.74 (1.94, 3.54) & 2.92 (2.37, 3.46) & 2.71 (2.44, 2.97) \\ 
        \textbf{ICL} & 2.57 (-3.02, 6.08) & 1.81 (-0.17, 3.78) & 2.32 (1.15, 3.50) & 2.16 (1.75, 2.57) \\ 
        \textbf{MALS} & 2.57 (-1.16, 6.08) & 1.81 (0.49, 3.12) & 2.32 (1.54, 3.11) & 2.16 (1.89, 2.43) \\ 
        \textbf{ME} & -1.93 (-15.00, 15.00) & -6.17 (-15.00, 15.00) & -0.24 (-15.00, 15.00) & -1.85 (-15.00, 15.00) \\ 
        \textbf{MQO\_MDH} & 5.10 (-15.00, 15.00) & 0.00 (-15.00, 15.00) & 7.16 (-15.00, 15.00) & 5.02 (-15.00, 15.00) \\ 
        \textbf{ODHC} & 1.81 (0.00, 3.86) & 2.47 (1.68, 3.27) & 2.68 (2.14, 3.23) & 2.48 (2.22, 2.74) \\ 
        \textbf{PCx\_ODX} & 0.17 (-15.00, 15.00) & 2.44 (-15.00, 15.00) & -0.04 (-15.00, 15.00) & 0.73 (-15.00, 15.00) \\ 
        \textbf{PDHC} & 0.04 (-4.56, 9.40) & 0.00 (-2.03, 2.03) & -0.00 (-1.12, 1.12) & -0.13 (-0.36, 0.10) \\ 
        \textbf{PEPCk\_PEPCx} & 0.28 (-15.00, 15.00) & -2.44 (-15.00, 15.00) & 1.59 (-15.00, 15.00) & 0.60 (-15.00, 15.00) \\ 
        \textbf{PFK} & 0.99 (0.16, 6.47) & 0.20 (0.16, 1.23) & 0.92 (0.21, 1.63) & 0.67 (0.35, 0.98) \\ 
        \textbf{PGD} & 2.38 (0.00, 15.00) & 0.00 (0.00, 3.14) & 2.25 (0.09, 4.41) & 1.51 (0.52, 2.50) \\ 
        \textbf{PGI} & 2.48 (0.09, 15.00) & 0.11 (0.09, 3.24) & 2.34 (0.18, 4.50) & 1.60 (0.61, 2.58) \\ 
        \textbf{PK\_PPS} & 1.24 (-15.00, 15.00) & 3.17 (-15.00, 15.00) & -0.20 (-15.00, 15.00) & 0.52 (-15.00, 15.00) \\ 
        \textbf{PTA} & 8.26 (7.33, 8.69) & 7.36 (7.33, 8.65) & 8.43 (7.33, 8.69) & 7.74 (7.33, 8.39) \\ 
        \textbf{RPE} & 1.53 (-0.05, 12.57) & -0.05 (-0.05, 2.04) & 1.45 (0.01, 2.90) & 0.96 (0.30, 1.63) \\ 
        \textbf{RPI} & 0.85 (0.04, 6.35) & 0.05 (0.04, 1.10) & 0.80 (0.08, 1.51) & 0.55 (0.22, 0.87) \\ 
        \textbf{SCS} & 1.74 (-0.07, 4.81) & 2.40 (1.21, 3.59) & 2.62 (1.80, 3.45) & 2.42 (2.03, 2.82) \\ 
        \textbf{SQO} & 4.38 (-0.27, 6.08) & 4.28 (3.02, 5.54) & 5.01 (4.00, 6.01) & 4.64 (4.21, 5.07) \\ 
        \textbf{TAL} & 0.79 (-0.00, 6.30) & -0.00 (-0.00, 1.04) & 0.75 (0.03, 1.47) & 0.50 (0.17, 0.83) \\ 
        \textbf{TKT1} & 0.79 (-0.00, 6.30) & -0.00 (-0.00, 1.04) & 0.75 (0.03, 1.47) & 0.50 (0.17, 0.83) \\ 
        \textbf{TKT2} & 0.74 (-0.05, 6.26) & -0.05 (-0.05, 1.00) & 0.71 (-0.02, 1.43) & 0.46 (0.13, 0.80) \\ 
        \textbf{TPI} & 0.99 (0.16, 6.47) & 0.20 (0.16, 1.23) & 0.92 (0.21, 1.63) & 0.67 (0.35, 0.98) \\ 
        \textbf{biomass} & 2.14 (1.80, 2.47) & 2.13 (1.81, 2.46) & 1.86 (1.53, 2.18) & 1.80 (1.58, 2.02) \\ 
        \textbf{growth} & 0.20 (0.17, 0.20) & 0.20 (0.17, 0.20) & 0.18 (0.17, 0.20) & 0.17 (0.17, 0.19) \\ 
        \textbf{production CO2} & 8.54 (6.68, 10.64) & 6.76 (6.68, 9.55) & 9.94 (7.14, 10.64) & 8.76 (7.31, 10.22) \\ 
        \textbf{production THF} & 0.02 (0.02, 0.02) & 0.02 (0.02, 0.02) & 0.02 (0.02, 0.02) & 0.02 (0.02, 0.02) \\ 
        \textbf{uptake} & 8.19 (6.87, 9.52) & 7.30 (6.01, 8.59) & 8.38 (7.08, 9.68) & 7.17 (6.52, 7.82) \\ 
        \bottomrule
    \label{sitab:TabS3}
\end{longtable}
\end{landscape}

\clearpage
\FloatBarrier

\begin{landscape}
\begin{longtable}{lcccc}
      \caption{\textbf{Pool size estimates obtained from independent analyses of the 1-, 2-, and U-\textsuperscript{13}C INST-ILE datasets, as well as from the joint evaluation of the 2- and U-\textsuperscript{13}C datasets.} Pool sizes are given in \SI{}{mmol.g_{X}^{-1}}. For each case, the upper value corresponds to the best-fit estimate, followed by the associated 95\% CIs in brackets below. During parameter estimation, pool sizes were constrained to the range between \SI{1e-4}{mmol.g_{X}^{-1}} and \SI{1e+1}{mmol.g_{X}^{-1}}. In the joint evaluation, selected metabolites were treated as ILE-specific parameters (i.e. AKG, ALA, ASP, CITR, CO2, DAP, DHAP, E4P, F6P, FBP, G6P, GA3P, GLU, HIS, HSER, LYS, ORN, PGP, PRA, R5P, RU5P, VAL, and XU5P), resulting in two separate estimates; the upper and lower values correspond to the 2-\textsuperscript{13}C and U-\textsuperscript{13}C sub-models, respectively.}
      \\

        \toprule
        \textbf{Metabolite} & 
        \textbf{1-\textsuperscript{13}C} & 
        \textbf{2-\textsuperscript{13}C} & 
        \textbf{U-\textsuperscript{13}C} & 
        \textbf{2-\& U-\textsuperscript{13}C} \\ 
        \midrule
        \endfirsthead

        \toprule
        \textbf{Metabolite} &
        \textbf{1-\textsuperscript{13}C} &
        \textbf{2-\textsuperscript{13}C} &
        \textbf{U-\textsuperscript{13}C} &
        \textbf{2-\& U-\textsuperscript{13}C} \\
        \midrule
        \endhead

        \bottomrule
        \endlastfoot
        
        \textbf{ACCOA} & \makecell{$1.09\cdot 10^{-4}$ \\ ($1.00\cdot 10^{-4}$, $1.00\cdot 10^{1}$)} & \makecell{$6.04\cdot 10^{-4}$ \\ ($1.00\cdot 10^{-4}$, $2.18\cdot 10^{0}$)} & \makecell{$1.67\cdot 10^{-3}$ \\ ($1.00\cdot 10^{-4}$, $1.09\cdot 10^{0}$)} & \makecell{$1.02\cdot 10^{-3}$ \\ ($1.00\cdot 10^{-4}$, $3.51\cdot 10^{-1}$)} \\ 
        \textbf{ACE} & \makecell{$1.11\cdot 10^{-4}$ \\ ($1.00\cdot 10^{-4}$, $1.00\cdot 10^{1}$)} & \makecell{$1.07\cdot 10^{-3}$ \\ ($1.00\cdot 10^{-4}$, $3.53\cdot 10^{-1}$)} & \makecell{$1.91\cdot 10^{-3}$ \\ ($1.00\cdot 10^{-4}$, $1.06\cdot 10^{0}$)} & \makecell{$1.31\cdot 10^{-3}$ \\ ($1.00\cdot 10^{-4}$, $6.81\cdot 10^{-2}$)} \\ 
        \textbf{AKG} & \makecell{$1.05\cdot 10^{-4}$ \\ ($1.00\cdot 10^{-4}$, $4.55\cdot 10^{-2}$)} & \makecell{$6.71\cdot 10^{-3}$ \\ ($1.00\cdot 10^{-4}$, $4.10\cdot 10^{-2}$)} & \makecell{$2.74\cdot 10^{-1}$ \\ ($2.16\cdot 10^{-1}$, $3.32\cdot 10^{-1}$)} & \makecell{$6.25\cdot 10^{-3}$ \\ ($1.00\cdot 10^{-4}$, $3.92\cdot 10^{-2}$) \\$2.18\cdot 10^{-1}$ \\ ($1.67\cdot 10^{-1}$, $2.69\cdot 10^{-1}$)} \\ 
        \textbf{ALA} & \makecell{$1.62\cdot 10^{-2}$ \\ ($9.75\cdot 10^{-3}$, $2.26\cdot 10^{-2}$)} & \makecell{$3.22\cdot 10^{-2}$ \\ ($2.62\cdot 10^{-2}$, $3.82\cdot 10^{-2}$)} & \makecell{$9.71\cdot 10^{-3}$ \\ ($2.56\cdot 10^{-3}$, $1.69\cdot 10^{-2}$)} & \makecell{$2.49\cdot 10^{-2}$ \\ ($1.94\cdot 10^{-2}$, $3.04\cdot 10^{-2}$) \\$9.10\cdot 10^{-3}$ \\ ($3.96\cdot 10^{-3}$, $1.42\cdot 10^{-2}$)} \\ 
        \textbf{AMP} & \makecell{$3.49\cdot 10^{-2}$ \\ ($1.00\cdot 10^{-4}$, $1.00\cdot 10^{1}$)} & \makecell{$3.11\cdot 10^{-2}$ \\ ($1.00\cdot 10^{-4}$, $1.00\cdot 10^{1}$)} & \makecell{$2.95\cdot 10^{-2}$ \\ ($1.00\cdot 10^{-4}$, $1.00\cdot 10^{1}$)} & \makecell{$3.18\cdot 10^{-2}$ \\ ($1.00\cdot 10^{-4}$, $1.00\cdot 10^{1}$)} \\ 
        \textbf{ARG} & \makecell{$3.72\cdot 10^{-2}$ \\ ($1.00\cdot 10^{-4}$, $1.00\cdot 10^{1}$)} & \makecell{$8.22\cdot 10^{-1}$ \\ ($4.56\cdot 10^{-1}$, $1.19\cdot 10^{0}$)} & \makecell{$1.88\cdot 10^{0}$ \\ ($1.00\cdot 10^{-4}$, $4.66\cdot 10^{0}$)} & \makecell{$1.74\cdot 10^{0}$ \\ ($3.04\cdot 10^{-1}$, $3.19\cdot 10^{0}$)} \\ 
        \textbf{ASP} & \makecell{$2.87\cdot 10^{-3}$ \\ ($1.03\cdot 10^{-3}$, $4.70\cdot 10^{-3}$)} & \makecell{$5.68\cdot 10^{-3}$ \\ ($3.09\cdot 10^{-3}$, $8.27\cdot 10^{-3}$)} & \makecell{$2.63\cdot 10^{-4}$ \\ ($1.00\cdot 10^{-4}$, $2.60\cdot 10^{-3}$)} & \makecell{$5.04\cdot 10^{-3}$ \\ ($3.31\cdot 10^{-3}$, $6.76\cdot 10^{-3}$) \\$3.26\cdot 10^{-4}$ \\ ($1.00\cdot 10^{-4}$, $1.49\cdot 10^{-3}$)} \\ 
        \textbf{CHOR} & \makecell{$3.35\cdot 10^{-1}$ \\ ($1.00\cdot 10^{-4}$, $1.00\cdot 10^{1}$)} & \makecell{$4.09\cdot 10^{-2}$ \\ ($1.00\cdot 10^{-4}$, $1.00\cdot 10^{1}$)} & \makecell{$1.07\cdot 10^{-1}$ \\ ($1.00\cdot 10^{-4}$, $1.00\cdot 10^{1}$)} & \makecell{$7.72\cdot 10^{-2}$ \\ ($1.00\cdot 10^{-4}$, $1.00\cdot 10^{1}$)} \\ 
        \textbf{CITR} & \makecell{$6.11\cdot 10^{-3}$ \\ ($5.06\cdot 10^{-3}$, $7.16\cdot 10^{-3}$)} & \makecell{$6.15\cdot 10^{-3}$ \\ ($5.16\cdot 10^{-3}$, $7.15\cdot 10^{-3}$)} & \makecell{$3.36\cdot 10^{-3}$ \\ ($2.29\cdot 10^{-3}$, $4.44\cdot 10^{-3}$)} & \makecell{$4.97\cdot 10^{-3}$ \\ ($4.04\cdot 10^{-3}$, $5.90\cdot 10^{-3}$) \\$3.41\cdot 10^{-3}$ \\ ($2.42\cdot 10^{-3}$, $4.40\cdot 10^{-3}$)} \\ 
        \textbf{CIT\_ICIT} & \makecell{$1.06\cdot 10^{-4}$ \\ ($1.00\cdot 10^{-4}$, $4.62\cdot 10^{-2}$)} & \makecell{$4.43\cdot 10^{-2}$ \\ ($1.49\cdot 10^{-2}$, $7.38\cdot 10^{-2}$)} & \makecell{$1.63\cdot 10^{-3}$ \\ ($1.00\cdot 10^{-4}$, $5.13\cdot 10^{-2}$)} & \makecell{$2.44\cdot 10^{-2}$ \\ ($3.22\cdot 10^{-3}$, $4.56\cdot 10^{-2}$)} \\ 
        \textbf{CO2} & \makecell{$5.27\cdot 10^{0}$ \\ ($1.00\cdot 10^{-4}$, $1.00\cdot 10^{1}$)} & \makecell{$1.52\cdot 10^{-2}$ \\ ($1.00\cdot 10^{-4}$, $2.19\cdot 10^{0}$)} & \makecell{$3.29\cdot 10^{0}$ \\ ($1.00\cdot 10^{-4}$, $9.83\cdot 10^{0}$)} & \makecell{$1.47\cdot 10^{-2}$ \\ ($1.00\cdot 10^{-4}$, $2.12\cdot 10^{0}$) \\$3.01\cdot 10^{0}$ \\ ($1.00\cdot 10^{-4}$, $1.00\cdot 10^{1}$)} \\ 
        \textbf{CYS} & \makecell{$5.47\cdot 10^{-2}$ \\ ($1.00\cdot 10^{-4}$, $1.00\cdot 10^{1}$)} & \makecell{$1.51\cdot 10^{-1}$ \\ ($1.00\cdot 10^{-4}$, $1.00\cdot 10^{1}$)} & \makecell{$1.05\cdot 10^{-1}$ \\ ($1.00\cdot 10^{-4}$, $1.00\cdot 10^{1}$)} & \makecell{$1.32\cdot 10^{-1}$ \\ ($1.00\cdot 10^{-4}$, $1.00\cdot 10^{1}$)} \\ 
        \textbf{DAP} & \makecell{$5.82\cdot 10^{-2}$ \\ ($4.64\cdot 10^{-2}$, $7.01\cdot 10^{-2}$)} & \makecell{$4.13\cdot 10^{-1}$ \\ ($3.09\cdot 10^{-1}$, $5.18\cdot 10^{-1}$)} & \makecell{$7.69\cdot 10^{-4}$ \\ ($1.00\cdot 10^{-4}$, $1.64\cdot 10^{-2}$)} & \makecell{$4.99\cdot 10^{-1}$ \\ ($3.24\cdot 10^{-1}$, $6.75\cdot 10^{-1}$) \\$6.33\cdot 10^{-4}$ \\ ($1.00\cdot 10^{-4}$, $1.39\cdot 10^{-2}$)} \\ 
        \textbf{DHAP} & \makecell{$1.04\cdot 10^{-2}$ \\ ($1.00\cdot 10^{-4}$, $1.00\cdot 10^{1}$)} & \makecell{$9.86\cdot 10^{0}$ \\ ($1.00\cdot 10^{-4}$, $1.00\cdot 10^{1}$)} & \makecell{$3.55\cdot 10^{-2}$ \\ ($1.00\cdot 10^{-4}$, $1.00\cdot 10^{1}$)} & \makecell{$8.82\cdot 10^{0}$ \\ ($1.00\cdot 10^{-4}$, $1.00\cdot 10^{1}$) \\$3.68\cdot 10^{-2}$ \\ ($1.00\cdot 10^{-4}$, $1.00\cdot 10^{1}$)} \\ 
        \textbf{E4P} & \makecell{$1.46\cdot 10^{-2}$ \\ ($1.00\cdot 10^{-4}$, $1.00\cdot 10^{1}$)} & \makecell{$5.15\cdot 10^{-3}$ \\ ($1.00\cdot 10^{-4}$, $1.00\cdot 10^{1}$)} & \makecell{$1.44\cdot 10^{-1}$ \\ ($1.00\cdot 10^{-4}$, $1.00\cdot 10^{1}$)} & \makecell{$5.80\cdot 10^{-3}$ \\ ($1.00\cdot 10^{-4}$, $1.00\cdot 10^{1}$) \\$1.59\cdot 10^{-1}$ \\ ($1.00\cdot 10^{-4}$, $1.00\cdot 10^{1}$)} \\ 
        \textbf{F6P} & \makecell{$5.23\cdot 10^{-3}$ \\ ($1.00\cdot 10^{-4}$, $1.00\cdot 10^{1}$)} & \makecell{$3.82\cdot 10^{-3}$ \\ ($1.00\cdot 10^{-4}$, $1.00\cdot 10^{1}$)} & \makecell{$8.72\cdot 10^{0}$ \\ ($1.00\cdot 10^{-4}$, $1.00\cdot 10^{1}$)} & \makecell{$3.66\cdot 10^{-3}$ \\ ($1.00\cdot 10^{-4}$, $1.00\cdot 10^{1}$) \\$8.61\cdot 10^{0}$ \\ ($1.00\cdot 10^{-4}$, $1.00\cdot 10^{1}$)} \\ 
        \textbf{FAICAR} & \makecell{$3.46\cdot 10^{-2}$ \\ ($1.00\cdot 10^{-4}$, $1.00\cdot 10^{1}$)} & \makecell{$2.94\cdot 10^{-2}$ \\ ($1.00\cdot 10^{-4}$, $1.00\cdot 10^{1}$)} & \makecell{$2.75\cdot 10^{-2}$ \\ ($1.00\cdot 10^{-4}$, $1.00\cdot 10^{1}$)} & \makecell{$2.72\cdot 10^{-2}$ \\ ($1.00\cdot 10^{-4}$, $1.00\cdot 10^{1}$)} \\ 
        \textbf{FBP} & \makecell{$2.81\cdot 10^{-3}$ \\ ($1.00\cdot 10^{-4}$, $1.00\cdot 10^{1}$)} & \makecell{$3.19\cdot 10^{-3}$ \\ ($1.00\cdot 10^{-4}$, $1.00\cdot 10^{1}$)} & \makecell{$4.60\cdot 10^{-2}$ \\ ($1.00\cdot 10^{-4}$, $1.00\cdot 10^{1}$)} & \makecell{$3.63\cdot 10^{-3}$ \\ ($1.00\cdot 10^{-4}$, $1.00\cdot 10^{1}$) \\$4.39\cdot 10^{-2}$ \\ ($1.00\cdot 10^{-4}$, $1.00\cdot 10^{1}$)} \\ 
        \textbf{FUM} & \makecell{$2.48\cdot 10^{-4}$ \\ ($1.00\cdot 10^{-4}$, $1.00\cdot 10^{1}$)} & \makecell{$5.93\cdot 10^{-4}$ \\ ($1.00\cdot 10^{-4}$, $4.04\cdot 10^{0}$)} & \makecell{$1.39\cdot 10^{-3}$ \\ ($1.00\cdot 10^{-4}$, $4.54\cdot 10^{0}$)} & \makecell{$1.03\cdot 10^{-3}$ \\ ($1.00\cdot 10^{-4}$, $1.45\cdot 10^{0}$)} \\ 
        \textbf{G6P} & \makecell{$7.06\cdot 10^{-3}$ \\ ($1.00\cdot 10^{-4}$, $1.00\cdot 10^{1}$)} & \makecell{$1.47\cdot 10^{-2}$ \\ ($1.00\cdot 10^{-4}$, $1.00\cdot 10^{1}$)} & \makecell{$4.61\cdot 10^{0}$ \\ ($1.00\cdot 10^{-4}$, $1.00\cdot 10^{1}$)} & \makecell{$1.75\cdot 10^{-2}$ \\ ($1.00\cdot 10^{-4}$, $1.00\cdot 10^{1}$) \\$4.11\cdot 10^{0}$ \\ ($1.00\cdot 10^{-4}$, $1.00\cdot 10^{1}$)} \\ 
        \textbf{GA3P} & \makecell{$2.39\cdot 10^{-3}$ \\ ($1.00\cdot 10^{-4}$, $1.00\cdot 10^{1}$)} & \makecell{$2.07\cdot 10^{-3}$ \\ ($1.00\cdot 10^{-4}$, $1.00\cdot 10^{1}$)} & \makecell{$3.30\cdot 10^{-2}$ \\ ($1.00\cdot 10^{-4}$, $1.00\cdot 10^{1}$)} & \makecell{$2.18\cdot 10^{-3}$ \\ ($1.00\cdot 10^{-4}$, $1.00\cdot 10^{1}$) \\$3.35\cdot 10^{-2}$ \\ ($1.00\cdot 10^{-4}$, $1.00\cdot 10^{1}$)} \\ 
        \textbf{GLN} & \makecell{$6.07\cdot 10^{-2}$ \\ ($5.09\cdot 10^{-2}$, $7.04\cdot 10^{-2}$)} & \makecell{$5.17\cdot 10^{-2}$ \\ ($4.37\cdot 10^{-2}$, $5.96\cdot 10^{-2}$)} & \makecell{$4.70\cdot 10^{-2}$ \\ ($3.88\cdot 10^{-2}$, $5.51\cdot 10^{-2}$)} & \makecell{$4.49\cdot 10^{-2}$ \\ ($3.94\cdot 10^{-2}$, $5.05\cdot 10^{-2}$)} \\ 
        \textbf{GLU} & \makecell{$1.48\cdot 10^{-1}$ \\ ($8.25\cdot 10^{-2}$, $2.13\cdot 10^{-1}$)} & \makecell{$8.46\cdot 10^{-2}$ \\ ($7.04\cdot 10^{-2}$, $9.88\cdot 10^{-2}$)} & \makecell{$1.10\cdot 10^{-2}$ \\ ($4.39\cdot 10^{-3}$, $1.76\cdot 10^{-2}$)} & \makecell{$8.99\cdot 10^{-2}$ \\ ($7.45\cdot 10^{-2}$, $1.05\cdot 10^{-1}$) \\$1.08\cdot 10^{-2}$ \\ ($6.84\cdot 10^{-4}$, $2.08\cdot 10^{-2}$)} \\ 
        \textbf{GLX} & \makecell{$1.63\cdot 10^{-2}$ \\ ($1.00\cdot 10^{-4}$, $7.24\cdot 10^{-2}$)} & \makecell{$8.09\cdot 10^{-2}$ \\ ($1.00\cdot 10^{-4}$, $2.10\cdot 10^{0}$)} & \makecell{$5.13\cdot 10^{-2}$ \\ ($4.96\cdot 10^{-3}$, $9.76\cdot 10^{-2}$)} & \makecell{$7.32\cdot 10^{-2}$ \\ ($1.00\cdot 10^{-4}$, $4.08\cdot 10^{-1}$)} \\ 
        \textbf{GLY} & \makecell{$1.01\cdot 10^{-4}$ \\ ($1.00\cdot 10^{-4}$, $1.21\cdot 10^{-3}$)} & \makecell{$9.99\cdot 10^{-2}$ \\ ($8.29\cdot 10^{-2}$, $1.17\cdot 10^{-1}$)} & \makecell{$6.96\cdot 10^{-2}$ \\ ($5.20\cdot 10^{-2}$, $8.72\cdot 10^{-2}$)} & \makecell{$8.72\cdot 10^{-2}$ \\ ($7.54\cdot 10^{-2}$, $9.89\cdot 10^{-2}$)} \\ 
        \textbf{HIS} & \makecell{$1.08\cdot 10^{-3}$ \\ ($1.00\cdot 10^{-4}$, $1.00\cdot 10^{1}$)} & \makecell{$2.90\cdot 10^{-4}$ \\ ($1.00\cdot 10^{-4}$, $1.00\cdot 10^{1}$)} & \makecell{$5.35\cdot 10^{-2}$ \\ ($1.00\cdot 10^{-4}$, $1.00\cdot 10^{1}$)} & \makecell{$2.94\cdot 10^{-4}$ \\ ($1.00\cdot 10^{-4}$, $3.90\cdot 10^{-1}$) \\$4.72\cdot 10^{-2}$ \\ ($1.00\cdot 10^{-4}$, $1.00\cdot 10^{1}$)} \\ 
        \textbf{HSER} & \makecell{$1.99\cdot 10^{-3}$ \\ ($1.64\cdot 10^{-3}$, $2.34\cdot 10^{-3}$)} & \makecell{$2.83\cdot 10^{-4}$ \\ ($1.00\cdot 10^{-4}$, $5.03\cdot 10^{-4}$)} & \makecell{$1.88\cdot 10^{-3}$ \\ ($1.48\cdot 10^{-3}$, $2.27\cdot 10^{-3}$)} & \makecell{$3.27\cdot 10^{-4}$ \\ ($1.33\cdot 10^{-4}$, $5.21\cdot 10^{-4}$) \\$1.49\cdot 10^{-3}$ \\ ($1.11\cdot 10^{-3}$, $1.86\cdot 10^{-3}$)} \\ 
        \textbf{I1} & \makecell{$2.76\cdot 10^{-2}$ \\ ($1.00\cdot 10^{-4}$, $1.00\cdot 10^{1}$)} & \makecell{$2.98\cdot 10^{-2}$ \\ ($1.00\cdot 10^{-4}$, $1.00\cdot 10^{1}$)} & \makecell{$3.32\cdot 10^{-2}$ \\ ($1.00\cdot 10^{-4}$, $1.00\cdot 10^{1}$)} & \makecell{$3.05\cdot 10^{-2}$ \\ ($1.00\cdot 10^{-4}$, $1.00\cdot 10^{1}$)} \\ 
        \textbf{I2} & \makecell{$2.88\cdot 10^{-2}$ \\ ($1.00\cdot 10^{-4}$, $1.00\cdot 10^{1}$)} & \makecell{$3.23\cdot 10^{-2}$ \\ ($1.00\cdot 10^{-4}$, $1.00\cdot 10^{1}$)} & \makecell{$3.63\cdot 10^{-2}$ \\ ($1.00\cdot 10^{-4}$, $1.00\cdot 10^{1}$)} & \makecell{$3.41\cdot 10^{-2}$ \\ ($1.00\cdot 10^{-4}$, $1.00\cdot 10^{1}$)} \\ 
        \textbf{IAP} & \makecell{$9.40\cdot 10^{-4}$ \\ ($1.00\cdot 10^{-4}$, $1.00\cdot 10^{1}$)} & \makecell{$2.94\cdot 10^{-1}$ \\ ($1.00\cdot 10^{-4}$, $3.91\cdot 10^{0}$)} & \makecell{$4.50\cdot 10^{-2}$ \\ ($1.00\cdot 10^{-4}$, $1.00\cdot 10^{1}$)} & \makecell{$1.53\cdot 10^{-1}$ \\ ($1.00\cdot 10^{-4}$, $4.96\cdot 10^{-1}$)} \\ 
        \textbf{ILE} & \makecell{$2.82\cdot 10^{-2}$ \\ ($2.21\cdot 10^{-2}$, $3.44\cdot 10^{-2}$)} & \makecell{$2.45\cdot 10^{-2}$ \\ ($1.00\cdot 10^{-4}$, $1.00\cdot 10^{1}$)} & \makecell{$3.58\cdot 10^{-2}$ \\ ($1.00\cdot 10^{-4}$, $1.00\cdot 10^{1}$)} & \makecell{$2.90\cdot 10^{-2}$ \\ ($1.00\cdot 10^{-4}$, $1.00\cdot 10^{1}$)} \\ 
        \textbf{IMP} & \makecell{$3.48\cdot 10^{-2}$ \\ ($1.00\cdot 10^{-4}$, $1.00\cdot 10^{1}$)} & \makecell{$3.04\cdot 10^{-2}$ \\ ($1.00\cdot 10^{-4}$, $1.00\cdot 10^{1}$)} & \makecell{$3.67\cdot 10^{-2}$ \\ ($1.00\cdot 10^{-4}$, $1.00\cdot 10^{1}$)} & \makecell{$3.03\cdot 10^{-2}$ \\ ($1.00\cdot 10^{-4}$, $1.00\cdot 10^{1}$)} \\ 
        \textbf{KIV} & \makecell{$3.61\cdot 10^{-3}$ \\ ($9.35\cdot 10^{-4}$, $6.28\cdot 10^{-3}$)} & \makecell{$2.47\cdot 10^{-3}$ \\ ($1.00\cdot 10^{-4}$, $2.73\cdot 10^{0}$)} & \makecell{$1.37\cdot 10^{-2}$ \\ ($1.00\cdot 10^{-4}$, $1.00\cdot 10^{1}$)} & \makecell{$7.62\cdot 10^{-3}$ \\ ($1.00\cdot 10^{-4}$, $1.66\cdot 10^{0}$)} \\ 
        \textbf{LYS} & \makecell{$1.04\cdot 10^{-4}$ \\ ($1.00\cdot 10^{-4}$, $4.92\cdot 10^{-4}$)} & \makecell{$2.45\cdot 10^{-4}$ \\ ($1.00\cdot 10^{-4}$, $2.55\cdot 10^{-3}$)} & \makecell{$5.82\cdot 10^{-1}$ \\ ($1.69\cdot 10^{-1}$, $9.95\cdot 10^{-1}$)} & \makecell{$2.32\cdot 10^{-4}$ \\ ($1.00\cdot 10^{-4}$, $3.39\cdot 10^{-3}$) \\$5.26\cdot 10^{-1}$ \\ ($1.89\cdot 10^{-1}$, $8.62\cdot 10^{-1}$)} \\ 
        \textbf{MAL} & \makecell{$1.76\cdot 10^{-4}$ \\ ($1.00\cdot 10^{-4}$, $1.00\cdot 10^{1}$)} & \makecell{$5.82\cdot 10^{-4}$ \\ ($1.00\cdot 10^{-4}$, $5.60\cdot 10^{0}$)} & \makecell{$1.33\cdot 10^{-3}$ \\ ($1.00\cdot 10^{-4}$, $6.59\cdot 10^{0}$)} & \makecell{$9.73\cdot 10^{-4}$ \\ ($1.00\cdot 10^{-4}$, $1.69\cdot 10^{0}$)} \\ 
        \textbf{NCLA} & \makecell{$8.95\cdot 10^{-2}$ \\ ($1.00\cdot 10^{-4}$, $1.00\cdot 10^{1}$)} & \makecell{$2.15\cdot 10^{-2}$ \\ ($1.00\cdot 10^{-4}$, $1.00\cdot 10^{1}$)} & \makecell{$2.98\cdot 10^{-2}$ \\ ($1.00\cdot 10^{-4}$, $1.00\cdot 10^{1}$)} & \makecell{$2.42\cdot 10^{-2}$ \\ ($1.00\cdot 10^{-4}$, $1.00\cdot 10^{1}$)} \\ 
        \textbf{OAA} & \makecell{$1.74\cdot 10^{-4}$ \\ ($1.00\cdot 10^{-4}$, $2.69\cdot 10^{0}$)} & \makecell{$9.28\cdot 10^{-4}$ \\ ($1.00\cdot 10^{-4}$, $2.05\cdot 10^{0}$)} & \makecell{$1.29\cdot 10^{-3}$ \\ ($1.00\cdot 10^{-4}$, $2.44\cdot 10^{0}$)} & \makecell{$9.69\cdot 10^{-4}$ \\ ($1.00\cdot 10^{-4}$, $2.70\cdot 10^{-1}$)} \\ 
        \textbf{ORN} & \makecell{$1.00\cdot 10^{-4}$ \\ ($1.00\cdot 10^{-4}$, $3.70\cdot 10^{-4}$)} & \makecell{$1.16\cdot 10^{-4}$ \\ ($1.00\cdot 10^{-4}$, $3.82\cdot 10^{-4}$)} & \makecell{$1.67\cdot 10^{-3}$ \\ ($9.04\cdot 10^{-4}$, $2.44\cdot 10^{-3}$)} & \makecell{$1.29\cdot 10^{-4}$ \\ ($1.00\cdot 10^{-4}$, $3.63\cdot 10^{-4}$) \\$1.67\cdot 10^{-3}$ \\ ($9.86\cdot 10^{-4}$, $2.35\cdot 10^{-3}$)} \\ 
        \textbf{ORO} & \makecell{$6.59\cdot 10^{-2}$ \\ ($1.00\cdot 10^{-4}$, $1.00\cdot 10^{1}$)} & \makecell{$2.66\cdot 10^{-2}$ \\ ($1.00\cdot 10^{-4}$, $1.00\cdot 10^{1}$)} & \makecell{$4.00\cdot 10^{-2}$ \\ ($1.00\cdot 10^{-4}$, $1.00\cdot 10^{1}$)} & \makecell{$3.00\cdot 10^{-2}$ \\ ($1.00\cdot 10^{-4}$, $1.00\cdot 10^{1}$)} \\ 
        \textbf{PEP} & \makecell{$1.88\cdot 10^{-4}$ \\ ($1.00\cdot 10^{-4}$, $1.00\cdot 10^{1}$)} & \makecell{$9.45\cdot 10^{-4}$ \\ ($1.00\cdot 10^{-4}$, $2.28\cdot 10^{0}$)} & \makecell{$7.55\cdot 10^{-4}$ \\ ($1.00\cdot 10^{-4}$, $2.97\cdot 10^{0}$)} & \makecell{$7.61\cdot 10^{-4}$ \\ ($1.00\cdot 10^{-4}$, $2.14\cdot 10^{-1}$)} \\ 
        \textbf{PGP} & \makecell{$8.07\cdot 10^{-4}$ \\ ($1.00\cdot 10^{-4}$, $1.00\cdot 10^{1}$)} & \makecell{$1.47\cdot 10^{-3}$ \\ ($1.00\cdot 10^{-4}$, $1.00\cdot 10^{1}$)} & \makecell{$6.97\cdot 10^{-2}$ \\ ($1.00\cdot 10^{-4}$, $6.69\cdot 10^{0}$)} & \makecell{$1.42\cdot 10^{-3}$ \\ ($1.00\cdot 10^{-4}$, $1.00\cdot 10^{1}$) \\$6.63\cdot 10^{-2}$ \\ ($1.00\cdot 10^{-4}$, $5.09\cdot 10^{0}$)} \\ 
        \textbf{PRA} & \makecell{$1.11\cdot 10^{-3}$ \\ ($1.00\cdot 10^{-4}$, $1.00\cdot 10^{1}$)} & \makecell{$2.90\cdot 10^{-4}$ \\ ($1.00\cdot 10^{-4}$, $1.00\cdot 10^{1}$)} & \makecell{$4.60\cdot 10^{-2}$ \\ ($1.00\cdot 10^{-4}$, $1.00\cdot 10^{1}$)} & \makecell{$3.36\cdot 10^{-4}$ \\ ($1.00\cdot 10^{-4}$, $3.74\cdot 10^{-1}$) \\$4.87\cdot 10^{-2}$ \\ ($1.00\cdot 10^{-4}$, $1.00\cdot 10^{1}$)} \\ 
        \textbf{PRE} & \makecell{$6.91\cdot 10^{-2}$ \\ ($1.00\cdot 10^{-4}$, $1.00\cdot 10^{1}$)} & \makecell{$1.67\cdot 10^{-2}$ \\ ($1.00\cdot 10^{-4}$, $1.00\cdot 10^{1}$)} & \makecell{$3.10\cdot 10^{-2}$ \\ ($1.00\cdot 10^{-4}$, $1.00\cdot 10^{1}$)} & \makecell{$2.43\cdot 10^{-2}$ \\ ($1.00\cdot 10^{-4}$, $1.00\cdot 10^{1}$)} \\ 
        \textbf{PRO} & \makecell{$3.94\cdot 10^{-2}$ \\ ($3.31\cdot 10^{-2}$, $4.57\cdot 10^{-2}$)} & \makecell{$2.51\cdot 10^{-2}$ \\ ($2.12\cdot 10^{-2}$, $2.89\cdot 10^{-2}$)} & \makecell{$2.18\cdot 10^{-2}$ \\ ($1.79\cdot 10^{-2}$, $2.56\cdot 10^{-2}$)} & \makecell{$2.19\cdot 10^{-2}$ \\ ($1.92\cdot 10^{-2}$, $2.46\cdot 10^{-2}$)} \\ 
        \textbf{PYR} & \makecell{$1.76\cdot 10^{-4}$ \\ ($1.00\cdot 10^{-4}$, $9.95\cdot 10^{0}$)} & \makecell{$9.03\cdot 10^{-4}$ \\ ($1.00\cdot 10^{-4}$, $5.41\cdot 10^{-1}$)} & \makecell{$8.20\cdot 10^{-4}$ \\ ($1.00\cdot 10^{-4}$, $9.76\cdot 10^{-1}$)} & \makecell{$7.63\cdot 10^{-4}$ \\ ($1.00\cdot 10^{-4}$, $1.94\cdot 10^{-1}$)} \\ 
        \textbf{R5P} & \makecell{$3.06\cdot 10^{-2}$ \\ ($1.00\cdot 10^{-4}$, $1.00\cdot 10^{1}$)} & \makecell{$3.92\cdot 10^{-3}$ \\ ($1.00\cdot 10^{-4}$, $1.00\cdot 10^{1}$)} & \makecell{$2.15\cdot 10^{-1}$ \\ ($1.00\cdot 10^{-4}$, $1.00\cdot 10^{1}$)} & \makecell{$4.05\cdot 10^{-3}$ \\ ($1.00\cdot 10^{-4}$, $1.00\cdot 10^{1}$) \\$1.82\cdot 10^{-1}$ \\ ($1.00\cdot 10^{-4}$, $1.00\cdot 10^{1}$)} \\ 
        \textbf{RU5P} & \makecell{$7.00\cdot 10^{-2}$ \\ ($1.00\cdot 10^{-4}$, $1.00\cdot 10^{1}$)} & \makecell{$3.07\cdot 10^{-3}$ \\ ($1.00\cdot 10^{-4}$, $1.00\cdot 10^{1}$)} & \makecell{$8.39\cdot 10^{-2}$ \\ ($1.00\cdot 10^{-4}$, $1.00\cdot 10^{1}$)} & \makecell{$3.59\cdot 10^{-3}$ \\ ($1.00\cdot 10^{-4}$, $1.00\cdot 10^{1}$) \\$9.14\cdot 10^{-2}$ \\ ($1.00\cdot 10^{-4}$, $1.00\cdot 10^{1}$)} \\ 
        \textbf{S7P} & \makecell{$1.68\cdot 10^{0}$ \\ ($1.00\cdot 10^{-4}$, $1.00\cdot 10^{1}$)} & \makecell{$2.56\cdot 10^{-2}$ \\ ($1.00\cdot 10^{-4}$, $1.00\cdot 10^{1}$)} & \makecell{$4.02\cdot 10^{-2}$ \\ ($1.00\cdot 10^{-4}$, $1.00\cdot 10^{1}$)} & \makecell{$4.46\cdot 10^{-2}$ \\ ($1.00\cdot 10^{-4}$, $1.00\cdot 10^{1}$)} \\ 
        \textbf{SER} & \makecell{$2.19\cdot 10^{-4}$ \\ ($1.00\cdot 10^{-4}$, $8.62\cdot 10^{-2}$)} & \makecell{$6.67\cdot 10^{-3}$ \\ ($7.14\cdot 10^{-4}$, $1.26\cdot 10^{-2}$)} & \makecell{$3.06\cdot 10^{-3}$ \\ ($1.00\cdot 10^{-4}$, $6.80\cdot 10^{-3}$)} & \makecell{$3.46\cdot 10^{-3}$ \\ ($1.87\cdot 10^{-3}$, $5.05\cdot 10^{-3}$)} \\ 
        \textbf{SHKM} & \makecell{$9.94\cdot 10^{-2}$ \\ ($1.00\cdot 10^{-4}$, $1.00\cdot 10^{1}$)} & \makecell{$1.68\cdot 10^{-2}$ \\ ($1.00\cdot 10^{-4}$, $1.00\cdot 10^{1}$)} & \makecell{$2.99\cdot 10^{-2}$ \\ ($1.00\cdot 10^{-4}$, $1.00\cdot 10^{1}$)} & \makecell{$2.35\cdot 10^{-2}$ \\ ($1.00\cdot 10^{-4}$, $1.00\cdot 10^{1}$)} \\ 
        \textbf{SUC} & \makecell{$2.94\cdot 10^{-4}$ \\ ($1.00\cdot 10^{-4}$, $1.00\cdot 10^{1}$)} & \makecell{$1.13\cdot 10^{-3}$ \\ ($1.00\cdot 10^{-4}$, $1.00\cdot 10^{1}$)} & \makecell{$1.54\cdot 10^{-3}$ \\ ($1.00\cdot 10^{-4}$, $8.04\cdot 10^{-1}$)} & \makecell{$1.08\cdot 10^{-3}$ \\ ($1.00\cdot 10^{-4}$, $2.16\cdot 10^{0}$)} \\ 
        \textbf{SUCCOA} & \makecell{$1.46\cdot 10^{-1}$ \\ ($1.00\cdot 10^{-4}$, $8.01\cdot 10^{-1}$)} & \makecell{$1.14\cdot 10^{-3}$ \\ ($1.00\cdot 10^{-4}$, $1.00\cdot 10^{1}$)} & \makecell{$1.68\cdot 10^{-3}$ \\ ($1.00\cdot 10^{-4}$, $7.43\cdot 10^{-1}$)} & \makecell{$1.20\cdot 10^{-3}$ \\ ($1.00\cdot 10^{-4}$, $2.13\cdot 10^{0}$)} \\ 
        \textbf{THF} & \makecell{$3.18\cdot 10^{-2}$ \\ ($1.00\cdot 10^{-4}$, $1.00\cdot 10^{1}$)} & \makecell{$2.48\cdot 10^{-2}$ \\ ($1.00\cdot 10^{-4}$, $1.00\cdot 10^{1}$)} & \makecell{$2.91\cdot 10^{-2}$ \\ ($1.00\cdot 10^{-4}$, $1.00\cdot 10^{1}$)} & \makecell{$2.86\cdot 10^{-2}$ \\ ($1.00\cdot 10^{-4}$, $1.00\cdot 10^{1}$)} \\ 
        \textbf{THR} & \makecell{$2.27\cdot 10^{-3}$ \\ ($1.81\cdot 10^{-3}$, $2.73\cdot 10^{-3}$)} & \makecell{$4.41\cdot 10^{-2}$ \\ ($3.66\cdot 10^{-2}$, $5.16\cdot 10^{-2}$)} & \makecell{$4.36\cdot 10^{-2}$ \\ ($3.59\cdot 10^{-2}$, $5.14\cdot 10^{-2}$)} & \makecell{$4.20\cdot 10^{-2}$ \\ ($3.67\cdot 10^{-2}$, $4.73\cdot 10^{-2}$)} \\ 
        \textbf{VAL} & \makecell{$2.54\cdot 10^{-4}$ \\ ($1.00\cdot 10^{-4}$, $1.61\cdot 10^{-3}$)} & \makecell{$2.40\cdot 10^{-3}$ \\ ($1.00\cdot 10^{-4}$, $3.47\cdot 10^{0}$)} & \makecell{$1.56\cdot 10^{-2}$ \\ ($1.00\cdot 10^{-4}$, $1.00\cdot 10^{1}$)} & \makecell{$2.19\cdot 10^{-3}$ \\ ($1.00\cdot 10^{-4}$, $1.00\cdot 10^{1}$) \\$1.61\cdot 10^{-2}$ \\ ($1.00\cdot 10^{-4}$, $1.00\cdot 10^{1}$)} \\ 
        \textbf{XU5P} & \makecell{$2.46\cdot 10^{0}$ \\ ($1.00\cdot 10^{-4}$, $1.00\cdot 10^{1}$)} & \makecell{$2.52\cdot 10^{-3}$ \\ ($1.00\cdot 10^{-4}$, $1.00\cdot 10^{1}$)} & \makecell{$3.47\cdot 10^{-2}$ \\ ($1.00\cdot 10^{-4}$, $1.00\cdot 10^{1}$)} & \makecell{$2.74\cdot 10^{-3}$ \\ ($1.00\cdot 10^{-4}$, $1.00\cdot 10^{1}$) \\$3.69\cdot 10^{-2}$ \\ ($1.00\cdot 10^{-4}$, $1.00\cdot 10^{1}$)} \\
    \label{sitab:TabS4}
\end{longtable}
\end{landscape}

\clearpage
\FloatBarrier

%%%%%%%%%%%%%%%%%%%%%%%%%%%%%%%%%%%%%%%%%%%%%
\begingroup{}
\renewcommand\refname{Supplementary References}
\bibliography{references}
\endgroup{}
%%%%%%%%%%%%%%%%%%%%%%%%%%%%%%%%%%%%%%%%%%%%%
\end{document}